\documentclass[aps,pre,preprint,groupedaddress,showpacs]{revtex4-1}
\usepackage[dvips]{graphicx}
\usepackage{textcomp}
\usepackage{amssymb}

\newcommand{\mc}{\multicolumn}

\begin{document}

\title{
\Large\bf Cubic fixed point in three dimensions:
  Monte Carlo simulations of the $\phi^4$ model on the lattice}

\author{Martin Hasenbusch}
\affiliation{
Institut f\"ur Theoretische Physik, Universit\"at Heidelberg,
Philosophenweg 19, 69120 Heidelberg, Germany}

\date{\today}
\begin{abstract}
We study the cubic fixed point for $N=3$ and $4$
by using finite-size scaling applied to data obtained from
Monte Carlo simulations of the $N$-component $\phi^4$ model on the simple 
cubic lattice.
We generalize the idea of improved models to a two-parameter family of models.
The two-parameter space is scanned for the point, where the amplitudes 
of the two leading 
corrections to scaling vanish. To this end, a dimensionless quantity is
introduced that monitors the breaking of the $O(N)$ invariance.
For $N=4$, we determine the correction exponents 
$\omega_1=0.763(24)$ and $\omega_2=0.082(5)$. In the case of $N=3$, we 
obtain $Y_4=0.0142(6)$ for the renormalization group exponent of the 
cubic perturbation at the $O(3)$-invariant fixed point, while the correction 
exponent $\omega_2=0.0133(8)$ at the cubic fixed point. Simulations close
to the improved point result in the estimates 
$\nu=0.7202(7)$ and $\eta=0.0371(2)$ of the critical exponents 
of the cubic fixed point for $N=4$. For $N=3$, at the cubic fixed point, 
the $O(3)$ symmetry is only mildly broken and the critical
exponents differ only by little from those of the $O(3)$-invariant fixed point.
We find 
$-0.00001  \lessapprox \eta_{cubic} - \eta_{O(3)} \lessapprox 0.00007$ and
$\nu_{cubic}-\nu_{O(3)} =-0.00061(10)$.
\end{abstract}

\maketitle
\section{Introduction}
The three-dimensional Heisenberg universality class is supposed to describe 
the critical behavior of isotropic magnets,
for example, the Curie transition in isotropic ferromagnets such as Ni and EuO, 
and of antiferromagnets such as RbMnF3 at the N\'eel transition point.
For a more detailed discussion, see for instance 
Sec. 5 of \cite{PeVi02} or the introduction of
Ref. \cite{ourHeisen}.
The Heisenberg universality class is characterized by an $O(3)$ symmetry 
of the order parameter.
Due to the crystal structure, in real systems, one expects that there are weak 
interactions 
that break the $O(3)$ symmetry and possess only cubic symmetry. 
Therefore, it is important to study the effect of such  perturbations
theoretically.
This has been done by using field-theoretic methods for five
decades now. In their seminal paper on the $\epsilon$ expansion \cite{WiFi72},
Wilson and Fisher discuss the problem to the leading order for $O(2)$ symmetry. 
Very soon the problem was taken on by various authors generalizing 
the calculation to $O(N)$ symmetry with arbitrary $N$ and extending the calculation to 
higher orders in the $\epsilon$ expansion. For example, in 1973, Aharony 
\cite{Aha73} performed the calculation to two-loop order.  Furthermore,
large-$N$ expansions around decoupled Ising systems were performed.
See Ref. \cite{Aha73Lett} and recently \cite{DBinder}.

It is beyond the scope of this paper to give a detailed account 
of the progress that has been made over the years. The development up to and 
the state of the art in 1999 is nicely summarized in Ref. \cite{Carmona}.
See also Refs. \cite{Varna00,Folk00}. At that time, the $\epsilon$ expansion
had been computed up to five loop \cite{Frohlinde} and the expansion in three 
dimensions fixed up to six loop  \cite{Carmona}.

\begin{figure}
\begin{center}
\includegraphics[width=14.5cm]{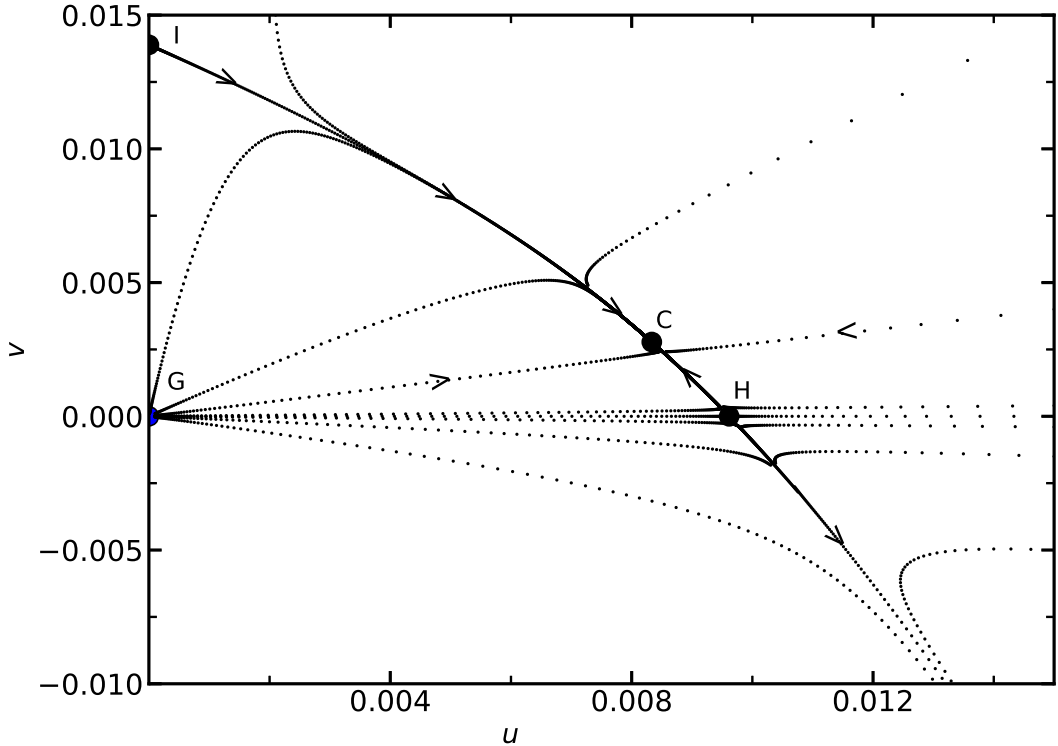}
\caption{\label{flowN5}
We have numerically integrated the one-loop flow equations~(\ref{RGeqCardy}) 
for $N=5$ and
$\epsilon=1$. The fixed points are given as solid circles.  The fixed 
points are labeled as G (Gaussian), I (decoupled Ising), H ($O(N)$-invariant
Heisenberg), and C (cubic). Selected RG trajectories are given by dotted lines.
Subsequent dots are separated by a scale factor of $b=2^{1/8}$. 
Hence, the larger the distance between the dots, the faster the flow.
The arrows indicate the direction of the flow.
}
\end{center}
\end{figure}

Here we like to summarize some basic facts to set the scene for our
numerical study. We follow the book \cite{Cardy}. Similar
discussions can be found in other reviews on the subject. 
The reduced Hamiltonian of the $\phi^4$ theory with two quartic couplings in 
the continuum is given by, [see, for example, Eq.~(5.66) of
Ref. \cite{Cardy}]
\begin{equation}
\label{ContHamil}
{\cal H} = \int \mbox{d}^d x  \left\{ 
\frac{1}{2} \sum_{i=1}^{N} [(\partial_{\mu}  \phi_i)^2 
+r \phi_i^{2}]  + \sum_{i,j=1}^{N} (u + v \delta_{ij})
\phi_i^2  \phi_j^2   \right\}  \;,
\end{equation}
where $\phi_i$ is a real number.
Note that for $d=4-\epsilon$, $\epsilon>0$ and $\epsilon$ small, these 
$\phi^4$ terms are the only relevant perturbations of the free (or Gaussian) 
theory.
For $v=0$ the system is $O(N)$-invariant. The question is, whether the term 
that breaks this invariance is relevant at the $O(N)$-invariant fixed point.
The qualitative picture already emerges from the leading-order calculation 
of the $\epsilon$ expansion. It can be obtained from the 
renormalization group (RG) flow on the critical surface
[Eqs.~(\ref{RGeqCardy})] 
taken from Eqs.~(5.67) and (5.68)  of Ref. 
\cite{Cardy}:
\begin{eqnarray}
\frac{\mbox{d} u}{\mbox{d} l} &=& \epsilon u - 8 (N + 8) u^2 - 48 u v + ... \;, \nonumber \\
\frac{\mbox{d} v}{\mbox{d} l} &=& \epsilon v -  96 u v -72 v^2 + ... \;,
\label{RGeqCardy}
\end{eqnarray}
where $l$ is the logarithm of a length scale. 
The set of differential equations~(\ref{RGeqCardy}) has four fixed points 
 \cite{Cardy}: 
\begin{itemize}
\item
Gaussian fixed point  $(u,v)=(0,0)$  ; 
\item
Decoupled Ising fixed point $(u,v)=(0,\epsilon/72)$ ; 
\item
$O(N)$-invariant fixed point $(u,v)=(\epsilon/(8 (N+8)),0)$  ;
\item
Cubic fixed point
$(u,v)=(\epsilon/(24 N),(N-4) \epsilon/(72 N))$ .
\end{itemize}
The Gaussian and the decoupled Ising fixed points are always unstable.
Whether the $O(N)$-invariant or the cubic fixed point is stable depends 
on $N$. For $N > N_c$, the cubic fixed point is stable, while for 
$N < N_c$ it is the $O(N)$-invariant fixed point. At one loop, setting 
$\epsilon=1$, $N_c=4$.  Analyzing higher orders in the 
$\epsilon$ expansion, $N_c \approx 3$ is obtained. The analysis of the 
five-loop $\epsilon$-expansion and the six-loop perturbative
series in three dimensions fixed gives $N_c$ slightly smaller than $3$  (see
Refs. \cite{Carmona,Varna00,Folk00} and references therein). 
Recently the $\epsilon$ expansion has been extended to six loop \cite{epsilon6}.
Analyzing the result, the authors find $N_c=2.915(3)$.
In Fig. \ref{flowN5} we give the flow obtained for $N=5$ and 
eqs.~(\ref{RGeqCardy}) for $\epsilon=1$.  The exact RG flows for $N=3$ and $4$
should show the same qualitative features. The $O(N)$-invariant fixed point 
has one stable direction, along the $u$-axis. The corresponding correction 
exponent is denoted by $\omega$. The RG exponent related with the unstable 
direction is denoted by $Y_4$, where the subscript refers to spin $l=4$.
At the cubic fixed point there are two stable directions characterized
by the correction exponents $\omega_2 < \omega_1$. The choice of the 
subscripts follows the literature.
RG trajectories starting with
$v<0$ run towards ever increasing $|v|$. Eventually $v$ reaches values, 
that give first-order transitions in mean field. Hence, one expects that for
any $v<0$, the transition is of first order.  A characteristic feature 
of the flow is that it collapses rather quickly on a single line, 
corresponding to the fact that $\omega$, $\omega_1$ $\gg$ $Y_4$, $\omega_2$.
For a recent reanalysis of the six-loop $\epsilon$ expansion and 
a discussion of the relevance in structural 
transitions in, for example, perovskites, see Ref. \cite{AharonyNeu}.

The result $N_c<3$ is supported by the fact that in a finite-size
scaling analysis of Monte Carlo data for the improved $\phi^4$ model on the 
simple cubic lattice the authors find $Y_4 = 0.013(4)$ for 
$N=3$ \cite{O234}.
The rigorous bound $Y_4 > 3-2.99056$ for $N=3$ was recently established 
by using the conformal bootstrap (CB) method \cite{CB_O3}. Note that
in the introduction of Ref. \cite{CB_O3} a nice summary of recent 
results obtained by different methods is given.

While it is established now that for $N=3$ the cubic fixed point 
is the stable one, highly accurate estimates of the critical exponents,
for example  the critical exponent $\nu$ of the correlation length, 
for the cubic fixed point are
missing. The accuracy of estimates obtained by using field theoretic methods 
does not allow to discriminate between the cubic and the $O(N)$-symmetric 
fixed point in the experimentally relevant case $N=3$.  Note that 
for the $O(3)$-invariant fixed point the estimates of critical exponents
obtained by 
Monte Carlo simulations of lattice models  (see, for example, Ref. \cite{myIco})
or the CB method \cite{CB_O3} are by one digit more accurate than those
obtained by field theoretic methods. For a more detailed discussion see 
Sec. \ref{summary} below.

In the present work, we provide accurate estimates of critical exponents
for the cubic fixed point for $N=3$ and $4$. 
To this end, we study a lattice version of the 
Hamiltonian~(\ref{ContHamil}) with two parameters. 
We generalize the idea of an improved model to a two-parameter  
model.
The idea to study improved models to
get better precision on universal quantities goes back to Refs.
\cite{ChFiNi,FiCh}. First studies of improved models 
using finite-size scaling (FSS) \cite{Barber}  
and Monte Carlo simulations applied to the three-dimensional Ising 
universality class are Refs. \cite{Bloete,Ballesteros,KlausStefano,myPhi4}.
For a discussion see, for example, Sec. 2.3 of the
review \cite{PeVi02}.
For the application to the three-dimensional Heisenberg universality class,
see Refs. \cite{O34,ourHeisen,O234,myIco}. By using finite-size scaling, 
one tunes the parameter of the reduced Hamiltonian such that the 
amplitude of the leading correction vanishes. Here in the case of the 
cubic fixed point, we are tuning two parameters to eliminate the 
amplitudes of the two leading corrections. Since one of the correction
exponents is much smaller than one, improving the reduced Hamiltonian 
turns out to be absolutely crucial to get reliable results for the 
critical exponents of the cubic fixed point.

In our simulations we consider $N=4$ in addition to $3$. 
It is not of direct experimental relevance.
However, here we expect that the cubic fixed point is well separated
from the $O(4)$-invariant
one, and therefore the conceptual points of our study can be more easily
demonstrated in this case.
Furthermore, our results allow to benchmark field-theoretic methods
that produce estimates for any value of $N$. 

The outline of the paper is the following: In the next section we define 
the model and the observables that we measure.
In Sec. \ref{algorithms} we discuss the Monte Carlo algorithms that are 
used for the simulation. In Sec. \ref{FSSsec} we summarize the theoretical 
predictions for the
FSS behavior of dimensionless quantities. In Sec. \ref{O4simulations} we
discuss the simulations and the analysis of the data for $N=4$.
For $N=3$ we first perform high statistics simulations 
at the $O(3)$-invariant point to improve the accuracy of the estimate of 
the exponent $Y_4$  (see Sec. \ref{O3exponents}).
Next, in Sec. \ref{O3exponents}, we discuss simulations for a finite 
perturbation with cubic symmetry. We locate the improved model
by analyzing the FSS behavior of dimensionless quantities. Estimates
of the critical exponents $\eta$ and $\nu$ are obtained by 
analyzing the FSS behavior of the magnetic susceptibility and the slope
of dimensionless quantities at criticality.
Finally, we summarize our results and compare them with estimates given in the 
literature.

\section{The model and observables}
\label{Model}
In our study we consider a discretized version of the continuum 
Hamiltonian~(\ref{ContHamil}) on a simple cubic lattice.
We extend the reduced Hamiltonian of the $\phi^4$ model  [see,
for example, Eq.~(1) of Ref. \cite{ourHeisen}]
by a term proportional to
\begin{equation}
 \sum_{a} Q_{4,a a a a}(\vec{\phi}\,) = \sum_{a} \phi_{x,a}^4
-  \frac{3}{N+2}
\left( \vec{\phi}_x^{\,2} \right)^2 \;,
\end{equation}
with cubic symmetry, breaking $O(N)$ invariance
\begin{equation}
\label{Hamiltonian}
 {\cal H}(\{\vec{\phi} \,\})= -\beta \sum_{<xy>} \vec{\phi}_x \cdot  \vec{\phi}_y
+ \sum_x \left [ \vec{\phi}_x^{\,2} + \lambda (\vec{\phi}_x^{\,2} -1)^2
+ \mu \left (\sum_{a} \phi_{x,a}^{\,4}
  -\frac{3}{N+2} \left( \vec{\phi}_x^{\,2} \right)^2 \right) \right]
\;,
\end{equation}
where 
$\vec{\phi}_x$ is a vector with $N$ real components.
The subscript $a$ denotes the components of the field and 
$\{\vec{\phi} \, \}$ is the collection of the fields at all sites  $x$.
We label the sites of the simple cubic lattice by
$x=(x_0,x_1,x_2)$, where $x_i \in \{1,2,\ldots,L_i\}$. Furthermore,
$<xy>$ denotes a pair of nearest neighbors on the lattice.
In our study, the linear lattice size $L=L_0=L_1=L_2$ is equal in all 
three directions throughout. We employ periodic boundary conditions.
The real 
numbers $\beta$, $\lambda$ and $\mu$ are the parameters of the 
model.
Note that
$Q_4$ is the traceless symmetric combination of four instances of 
the field [see, for example, Eq.~(7) of Ref. \cite{O234}] 
\begin{eqnarray}
Q_{4, abcd}(\vec{\Phi}\,) &=& \Phi_a \Phi_b \Phi_c \Phi_d
\nonumber \\
&-& {1\over N+4} \vec{\Phi}^{\,2} \left(
        \delta_{ab} \Phi_c \Phi_d + \delta_{ac} \Phi_b \Phi_d +
        \delta_{ad} \Phi_b \Phi_c + \delta_{bc} \Phi_a \Phi_d +
        \delta_{bd} \Phi_a \Phi_c + \delta_{cd} \Phi_a \Phi_b \right)
\nonumber \\
    &+& {1\over (N+2)(N+4)} (\vec{\Phi}^{\,2})^2 \left(
         \delta_{ab} \delta_{cd} + \delta_{ac} \delta_{bd} +
         \delta_{ad} \delta_{bc} \right)  \;.
\label{spin4}
\end{eqnarray}
The expectation value of $\sum_{a} Q_{4, a a a a}(\vec{\phi} \,)$
vanishes for an $O(N)$-symmetric distribution of $\vec{\phi}$.
A small perturbation of the $O(N)$-symmetric system, $\mu=0$, 
only affects scaling fields with the symmetries corresponding 
to the cubic symmetry of the perturbation. 
See also Eq.~(5) of Ref. \cite{CB_O3} and the accompanying discussion.

Note that in the Hamiltonian~(\ref{Hamiltonian}) the components of the field 
decouple for $\lambda -\frac{3}{N+2} \mu=0$. Since the term 
$\sum_x \vec{\phi}_x^{\,2}$ has the factor $(1-2 \lambda)$ and 
$\sum_x \sum_a \phi_{x,a}^{4}$ the factor $\mu = \frac{N+2}{3} \lambda$
in front, a rescaling of the field $\phi_x$ is needed to match with the 
Hamiltonian 
\begin{equation}
\label{HamiltonianI}
 {\cal H}(\{\phi\})= -\tilde \beta \sum_{<xy>} \phi_x \phi_y
+ \sum_x \left [ \phi_x^{2} + \tilde \lambda (\phi_x^{2} -1)^2
   \right]
\;,
\end{equation}
considered for example in ref. \cite{myPhi4}, where $\phi_x$ is a real 
number. We arrive at the equations
\begin{equation}
 (1-2 \lambda)  = (1-2 \tilde \lambda) \; c \;\; ,\;\;\;\; 
\frac{N+2}{3} \lambda = \tilde \lambda \; c^2
\end{equation}
and hence
\begin{equation}
\frac{6}{N+2} \tilde \lambda \; c^2 +  (1-2 \tilde \lambda) \; c -1 =0 
\end{equation}
with the solutions
\begin{equation}
c =  \frac{-(1-2 \tilde \lambda) \pm \sqrt{(1-2 \tilde \lambda)^2 
 +\frac{24}{N+2} \tilde \lambda}}{\frac{12}{N+2} \tilde \lambda} \;\;,
\end{equation}
where we take the positive solution. Plugging in $\tilde \lambda^*\approx 1.1$
\cite{myPhi4} we arrive at $c=1.43647586...$ for $N=3$. Note that
$\tilde \lambda^*$ denotes the value of $\tilde \lambda$, where leading 
corrections to scaling vanish. Hence we get for the improved decoupled
model
$\lambda^*=1.361885...$, $\mu^* = \frac{N+2}{3} \lambda^* =2.269809...$, and 
$\tilde \beta_c =0.3750966(4)$ at $\tilde \lambda=1.1$ translates to
$\beta_c =c \tilde \beta_c = 0.5388172 ...\;$.  For $N=4$ we get
$\lambda^*=1.486347...$, $\mu^* = 2.972695 ...$, and $\beta_c = 0.616626 ...\;$.

\subsection{The observables and dimensionless quantities}
\label{observables}
Dimensionless quantities or phenomenological couplings play a central
role in finite size scaling. 
Similar to the study of $O(N)$-invariant models we study
the Binder cumulant $U_4$, the ratio of partition functions $Z_a/Z_p$ and
the second moment correlation length over the linear lattice size
$\xi_{2nd}/L$. Let us briefly recall the definitions of the observables
and dimensionless quantities that we measure.

The energy of a given field configuration is defined as
\begin{equation}
\label{energy}
 E=  \sum_{<xy>}  \vec{\phi}_x  \cdot \vec{\phi}_y \;\;.
\end{equation}

The magnetic susceptibility $\chi$ and the second moment correlation length
$\xi_{2nd}$ are defined as
\begin{equation}
\label{suscept}
\chi  \equiv  \frac{1}{V}
\left\langle \Big(\sum_x \vec{\phi}_x \Big)^2 \right\rangle \;\;,
\end{equation}
where $V=L^3$ and
\begin{equation}
\xi_{2nd}  \equiv  \sqrt{\frac{\chi/F-1}{4 \sin^2 \pi/L}} \;\;,
\end{equation}
where
\begin{equation}
F  \equiv  \frac{1}{V} \left \langle
\Big|\sum_x \exp\left(i \frac{2 \pi x_k}{L} \right)
        \vec{\phi}_x \Big|^2
\right \rangle
\end{equation}
is the Fourier transform of the correlation function at the lowest
non-zero momentum. In our simulations, we have measured $F$ for the three
directions $k=0,1,2$ and have averaged these three results.

The Binder cumulant $U_4$ is given by
\begin{equation}
U_{4} \equiv \frac{\langle (\vec{m}^{2})^2 \rangle}{\langle \vec{m}^2\rangle^2},
\end{equation}
where $\vec{m} = \frac{1}{V} \, \sum_x \vec{\phi}_x$ is the
magnetization of a given field configuration. We also consider the ratio
$Z_a/Z_p$ of
the partition function $Z_a$ of a system with anti-periodic boundary
conditions in one of the three directions and the partition function
$Z_p$ of a system with periodic boundary conditions in all directions.
This quantity is computed by using the cluster algorithm.
For a discussion see Appendix A 2 of ref. \cite{XY1}.

In order to detect the effect of the cubic anisotropy we
study
\begin{equation}
\label{UCdef}
U_C = \frac{\langle \sum_a Q_{4,aaaa}(\vec{m}) \rangle}
     {\langle \vec{m}^{\,2} \rangle^2}  \; .
\end{equation}
In the following we shall refer to the RG-invariant
quantities $U_C$, $U_{4}$, $Z_a/Z_p$, and $\xi_{2nd}/L$ by
using the symbol $R$.

In our analysis we need the observables as a function of $\beta$ in
some neighborhood of the simulation point $\beta_s$. To this end we have
computed the coefficients of the Taylor expansion of the observables
up to the third order.
For example the first derivative of the expectation value
$\langle A \rangle$ of an observable $A$ is given by
\begin{equation}
\frac{\partial \langle A \rangle}{\partial \beta} = \langle A E \rangle
- \langle A \rangle \langle E \rangle \;\;.
\end{equation}

In the case of decoupled systems, $\lambda -\frac{N+2}{3} \mu=0$, we can express
the dimensionless quantities introduced above in terms of their Ising 
counterparts.
For example 
\begin{equation}
U_C = \frac{N-1}{N (N+2)} (U_{4,Ising} - 3) \;\;.
\end{equation}
Hence, we get for the fixed point value, which is indicated by $^*$
\begin{equation}
U_{C,DI}^*= (1.60359(4) -3)  \frac{N-1}{N (N+2)} = - 1.39641(4) 
\frac{N-1}{N (N+2)}
\end{equation}
using the result of \cite{myIso} for  $U_{4,Ising}^*$. Furthermore,
$(Z_a/Z_p)^*_{DI} = ((Z_a/Z_p)^*_{Ising})^N$, $U_{4,DI}^* = 
\frac{1}{N} U_{4,Ising}^* + \frac{N-1}{N}$, and 
$(\xi_{2nd}/L)^*_{DI} = (\xi_{2nd}/L)^*_{Ising}$, where the subscript
$DI$ indicates the decoupled Ising fixed point.

\section{Monte Carlo algorithm}
\label{algorithms}
In previous work (see, for example, Refs. \cite{O34,O234,myON})
we have simulated the 
$O(N)$-invariant $\phi^4$ model on the simple cubic lattice. Here, for $N=4$,
we use the algorithm and C-code of ref. \cite{myON} with minor 
modifications, to take into account the term proportional to $\mu$ in the
reduced Hamiltonian~(\ref{Hamiltonian}).  For $N=3$, we have implemented 
the algorithm by using AVX intrinsics to speed up the simulation.

The algorithm used in ref.  \cite{myON} is a hybrid of:
\begin{itemize}
\item wall cluster algorithm \cite{KlausStefano};
\item local Metropolis update;
\item local over-relaxation update;
\item global rotation of the field.
\end{itemize}
In the case of the wall cluster update, in ref. \cite{myON}, we performed
the update for technical reasons componentwise. This means that in a given
update step
only the sign of a single component of the field might change. This way, the 
value of $\sum_{a} Q_{4, a a a a}(\vec{\phi}_x)$ remains unchanged.
Hence, we can take this part of the C-program from Ref. \cite{myON}
without any change.  Also the measurement of $Z_a/Z_p$, which is integrated 
into the wall cluster update, is reused without change.

In the local Metropolis algorithm, we generate a proposal by
\begin{equation}
 \phi_{x,i}' =  \phi_{x,i} + s (r_i -0.5)
\end{equation}
for each component $i$ of the field at the site $x$. $r_i$ is a uniformly 
distributed random number in $[0,1)$ and the step size $s$ is tuned such
that the acceptance rate is roughly $50 \%$. Note that for each component
a random number $r_i$ is taken.
We use the  acceptance probability
\begin{equation}
\label{Pacc}
 P_{acc} = \mbox{min}\left[1,\exp(-H(\{\vec{\phi}\,\}') + H(\{\vec{\phi}\,\})\right]  \;.
\end{equation}
The only change compared with the program of Ref. \cite{myON} is
that we have to take into account the term 
$\mu \sum_{a} Q_{4, a a a a}(\vec{\phi}_x)$, when computing
$\Delta H(\{\vec{\phi}\,\}',\{\vec{\phi}\,\}) = H(\{\vec{\phi}\,\}') -H(\{\vec{\phi}\,\})$. 

We have implemented over-relaxation updates
\begin{equation}
 \vec{\phi}_x^{\;\;'} =
 2 \frac{\vec{\Phi}_x \cdot \vec{\phi}_x}{\vec{\Phi}_x^2} \vec{\Phi}_x
- \vec{\phi}_x \;\;,
 \end{equation}
 where
\begin{equation}
\vec{\Phi}_x = \sum_{y.nn.x}  \vec{\phi}_y \;\;,
\end{equation}
where $\sum_{y.nn.x}$ is the sum over all nearest neighbors $y$ of the
site $x$.  In the case of the $O(N)$-invariant $\phi^4$ model
this update does not change the value of the Hamiltonian and
therefore no accept/reject step is needed.
Here, the value of the term $\mu \sum_{a} Q_{4, a a a a}(\vec{\phi}_x)$
changes under the update, 
which has to be taken into account in an accept/reject step~(\ref{Pacc}). 
This update has no parameter which can be tuned. The acceptance rate depends
on the parameters of the model. In particular, the larger $\mu$, the smaller 
the acceptance rate. It turns out that for the range of $\mu$ studied here 
the acceptance rate is reasonably high.  For example, for $N=4$, 
for $(\lambda,\mu)=(7,2.64)$, which is close to the improved point
$(\lambda,\mu)^*$, we get an acceptance rate of about $0.77$ at
the critical point, with little dependence on the lattice size.
In the case of $N=3$, the values of $\mu$ that we simulated at are smaller
and hence the acceptance rates are even larger.

In Ref.  \cite{myON} we use global rotations of the field to 
compensate for the fact that the cluster update has preferred directions. 
The global rotation changes the value of the new  term 
$\sum_x \mu \sum_{a} Q_{4, a a a a}(\vec{\phi})$. Hence, 
an accept/reject step has to be introduced. In addition, we introduced 
a step size for the global rotation, which is tuned such that the 
acceptance rate is very roughly $1/2$.  For simplicity we did not 
perform a general $O(N)$ rotation, but used a rotation among two of the 
components.  
It turned out that these global rotations are useful only for small $\mu$
and/or small linear lattice sizes $L$. In particular for $\mu$ of the 
order of $\mu^*$, the reduction  in the autocorrelation times, for reasonable
lattice sizes, does not pay off for the computational costs of the rotation.
Therefore, eventually, we skipped this component of the update. 
Unfortunately this leaves us with the potential problem that the cluster
update discussed above only changes the sign of a given component of the field.

In fact, for lattice sizes $L \gtrapprox 32$ it turned out to be advantageous
to add cluster updates that exchange two components of the field 
\begin{equation}
\label{nosing}
\phi_{x,j}' = \phi_{x,i}  \;\;, \; \phi_{x,i}' = \phi_{x,j}
\end{equation}
for $i \ne j$, while the other components stay unchanged.
Note that this update leaves the term $\sum_{a} Q_{4,a a a a}(\vec{\phi}_{x})$
unchanged. The update can be written as a reflection
\begin{equation}
\vec{\phi}_x' = \vec{\phi}_x -2 (\vec{r} \cdot \vec{\phi}_x) \vec{r}
\end{equation}
with $r_i=2^{-1/2} $, $r_j=-2^{-1/2}$, and $r_k=0$ for $k \ne i,j$.
The cluster update can also be performed with an additional change of the sign:
\begin{equation}
\label{chsing}
\phi_{x,j}' = - \phi_{x,i}  \;\;, \; \phi_{x,i}' = - \phi_{x,j}
\end{equation}
for $i \ne j$, while the other components stay unchanged.
For simplicity, 
we have implemented this update as single-cluster update \cite{Wolff}.
With probability $1/2$ we took either eq.~(\ref{nosing}) or 
eq.~(\ref{chsing}) for a given cluster update.

During the major part of the simulations, we did not measure autocorrelation
times, since we performed binning of the data at run time.
In preliminary simulations, where we performed of the order of $10^6$
update cycles, we stored all measurement. In the analysis we computed
integrated autocorrelation times for a selection of 
observables that we studied. 

In the case of $N=4$ an update and measurement cycle is given by the following
C-code: \\
\verb|over(); rotate(); metro(); for(ic=0;ic<N;ic++) wall_0(ic); measure();| \\
\verb|over(); rotate(); metro(); for(ic=0;ic<N;ic++) wall_1(ic); measure();| \\
\verb|over(); rotate(); metro(); for(ic=0;ic<N;ic++) wall_2(ic); measure();| \\
Here \verb+over()+ is a sweep with the over-relaxation update over the
lattice. \verb+rotate()+ is the global rotation of the field. For larger 
lattices, we have skipped the rotation. \verb+metro()+ is a sweep with the 
Metropolis update discussed above, followed by an over-relaxation update
at the same site.
\verb+wall_k(ic)+ is a wall-cluster update with a plane perpendicular to the
\verb+k+-axis. The component \verb+ic+ of the field is updated.
\verb+measure()+ contains the evaluation of the observables discussed 
above. 

In the most recent version of the program, a sequence of single-cluster 
updates replaces \verb+rotate()+. The sequence 
is given by \\
\verb| for(j=0;j<L/8;j++) for(l=0;l<N-1;l++) for(k=l+1;l<N;l++) single(l,k);| \\
where  \verb+ single(l,k) + is a single-cluster update, exchanging the 
components \verb+l+ and  \verb+k+ for the sites within the cluster.

In Table \ref{auto_N4} we give integrated autocorrelation times for 
the energy, the magnetic susceptibility $\chi$, and $Q_4(\vec{m})$ for 
$(\lambda,\mu)=(7,2.64)$, which is close to $(\lambda,\mu)^*$, at 
$\beta=0.86407506$, which is our estimate of $\beta_c$.  We truncated 
the summation of the integrated autocorrelation function at 
$t_{max}=6 \tau_{int,E}$ for all quantities considered. Throughout 
we performed $10^6$ measurements. Note that adding the single-cluster updates
increases the CPU time needed for one update cycle by about 40 $\%$.
Hence already for $L=32$ we see an advantage for adding the 
single-cluster updates.

\begin{table}
\caption{\sl \label{auto_N4}
Estimates of the integrated autocorrelation time $\tau_{int}$ of the
energy $E$, the magnetic susceptibility $\chi$, and $Q_4(\vec{m})$ for
$N=4$, at $(\lambda,\mu)=(7,2.64)$ and $\beta=0.86407506$. ''single''
and ''no single'' refer to simulations that have component exchanging
single-cluster updates or not. For a discussion see the text.
}
\begin{center}
\begin{tabular}{rccccc}
\hline
$L$  &   type  &$t_{max}$&  $\tau_{int,E}$    & $\tau_{int,\chi}$ & $\tau_{int,Q_4}$ \\
\hline
16 & no  single  & 30  & 4.76(5)  & 3.58(3)& 2.37(3) \\
16 &    single   & 21  & 3.38(3)  & 2.30(2)& 1.57(1) \\
32 & no  single  & 38  & 6.16(7)  & 4.48(5)& 3.34(4) \\
32 &    single   & 26  & 4.20(4)  & 2.40(2)& 1.63(2) \\
64 &  no  single & 50  & 8.29(11) & 5.65(7)& 4.62(7) \\
64 &  single     & 31  & 5.09(5)  & 2.44(2)& 1.50(2) \\
128&  no  single & 68  &11.13(17) & 7.24(11)& 6.11(12) \\
128&  single     & 38  &6.35(7)   & 2.51(3) & 1.41(2) \\
\hline
\end{tabular}
\end{center}
\end{table}

In the case of $N=3$, we implemented the algorithm by using 
the AVX instruction set of x86 CPUs. These were accessed by using
AVX intrinsics. AVX instructions act on several variables that are packed 
into 256 bit units in parallel.
In particular we used \verb+__m256d+ variables to store 4 double
precision floating point numbers. We employed a trivial parallelization,
simulating four systems in parallel. Each of the floating point numbers
in a \verb+__m256d+ variable 
is associated with one of the four systems that is simulated. 
This way, we could speed up the local updates and the measurement of 
the observables by a factor somewhat larger than two. 
To this end we reused the random number $r^{(0)}$ that is uniformly
distributed in $[0,1)$ by
\begin{equation}
 r^{(j)} = \mbox{frac}(r^{(0)} +j/4) \;,
\end{equation}
where $j = 0$, $1$, $2$, or $3$ and frac is the fractional part
of a real number. A discussion on the reuse of random 
numbers is given in Appendix A of Ref. \cite{ourDynamic}.

In the case of the cluster algorithm we found no efficient use of the 
parallel execution using AVX instruction. Therefore, we go through the 
four copies of the field sequentially. Here, the data layout is a small
obstacle. Therefore, the overall gain obtained by using the parallelization
as discussed above is at the level of about $20 \%$. 

Since the overall gain is rather modest compared with a plain C 
implementation, 
we abstain from a detailed discussion of the implementation. We experimented
with the composition of the update cycle. It turns out that the dependence 
of the efficiency on the precise composition is rather flat. Similar to  
$N=4$, it is clearly advantageous to add cluster updates that exchange 
components of the field. The update and measurement cycles used 
in most of our simulations are similar to those discussed above for $N=4$. 
Motivated by the speedup of the local updates by the AVX implementation, 
however, the relative number of local over-relaxation updates compared with the 
cluster updates is increased.

\section{Finite-size scaling}
\label{FSSsec}
In this section we recall the theoretical basis of the FSS analysis of
dimensionless quantities. 
In particular, we consider the ratio of partition functions $Z_a/Z_p$,
the second moment correlation length over the linear lattice size
$\xi_{2nd}/L$, the Binder cumulant $U_4$, and the quantity $U_C$ that
quantifies the violation of the $O(N)$ symmetry.
First we consider the neighborhood of a
single fixed point, being well separated from other fixed points. 
In previous work, we discussed the case of a single correction with 
a correction exponent $\omega$ being clearly smaller than two. 
Here we discuss the case of two such corrections with the exponents
$\omega_2 < \omega_1< 2$, which is the case for the cubic fixed point. 

This turns out to be sufficient for the analysis of the cubic fixed 
point for $N=4$. However, for $N=3$ it is desirable  to
extend the discussion to the neighborhood of two fixed points
that are close to each other. 

\subsection{Dimensionless quantities in the neighborhood of a fixed point}
Dimensionless quantities $R_i$, for a given geometry, are functions of the 
lattice size $L$ and the parameters $\beta$, $\lambda$, and $\mu$ of the
reduced Hamiltonian. 
Throughout, we consider a vanishing external field $h=0$. In the neighborhood
of a critical point, 
we might also write them as a function of the nonlinear scaling fields $u_j$ 
and the linear lattice size $L$:
\begin{equation}
\label{Rscaling}
 R_i(\beta,\lambda,\mu,L) = R_i(u_t L^{y_t},u_3 L^{y_3}, u_4 L^{y_4}, \{u_j L^{y_j} \}) \;,
\end{equation}
where we identify $y_3=-\omega_2$ and $y_4 = - \omega_1$ in the case of the
cubic fixed point. $y_t=1/\nu$ is the thermal RG exponent. 
Note that $y_3 > y_4 > -1$, while we expect, similar as for
the $O(N)$-invariant models, $y_j \lessapprox -2$ for $j>4$. 
The non-linear scaling fields can be written as (see, for example, Ref. 
\cite{PeVi02}, sec. 1.5.7)
\begin{equation}
u_t = t +  g_{11}(\lambda,\mu) t^2 + O(t^3) \; ,
\end{equation}
where $t=\beta-\beta_c$ is the reduced temperature, 
and $\beta_c$ is the inverse 
critical temperature. For simplicity, in the definition of $t$, we have skipped
the normalization $1/\beta_c$ and took the opposite sign as usual.
The irrelevant scaling fields are
\begin{equation}
u_3 = g_{13}(\lambda,\mu) + g_{23}(\lambda,\mu)  t + O(t^2)
\end{equation}
and
\begin{equation}
u_4 = g_{14}(\lambda,\mu) + g_{24}(\lambda,\mu)  t + O(t^2) \; .
\end{equation}

Now let us expand $R_i$ on the right-hand side of eq.~(\ref{Rscaling}) around
the fixed point $u_j L^{y_j}=0$:
\begin{equation}
R_i(\beta,\lambda,\mu,L) = R_i^* + \sum_{j} r_{i,j} u_j L^{y_j} + 
\frac{1}{2}  \sum_{j,k} r_{i,j,k} u_j u_k L^{y_j+y_k} + ... \;,
\end{equation}
where $j = t, 3, 4, ...$ and
\begin{equation}
r_{i,j} = \frac{\partial R_i}{\partial (u_j L^{y_j})}
\end{equation}
and
\begin{equation}
r_{i,j,k} = \frac{\partial^2 R_i}{\partial (u_j L^{y_j}) \partial (u_k L^{y_k})}
\end{equation}
at the fixed point. Now putting in the expressions for the scaling 
fields $u_j$ we arrive at
\begin{equation}
\label{startingR}
R_i(\beta_c,\lambda,\mu,L) = R_i^* + \sum_{j \ge 3} r_{i,j} g_{1j}(\lambda,\mu) L^{y_j}
+ ... \; 
\end{equation}
at the critical temperature. Eq.~(\ref{startingR}) is the basis for the 
Ans\"atze used to fit our data. Note that we have simulated at 
$\beta_s \approx \beta_c$. In addition to the value of 
$R_i(\beta_s,\lambda,\mu,L)$, we determine the Taylor coefficients of the 
expansion of $R_i$ in $(\beta-\beta_s)$ up to the third order. In our
fits, we keep $R_i(\beta_s,\lambda,\mu,L)$ on the left side of the equation,
and bring the terms proportional to $(\beta_c-\beta_s)^{\alpha}$ for 
$\alpha=1$, $2$, and $3$ to the left. Furthermore, we ignore the statistical 
error of the Taylor coefficients.  This way, we can treat $\beta_c$ as a 
free parameter in the fit.

In order to arrive at an Ansatz that can be used in a fit, we have to truncate
eq.~(\ref{startingR}). After a few numerical experiments we took
\begin{eqnarray}
\label{RGansatz1}
 R_i(\beta_c,\lambda,\mu,L) &=& R_i^* + r_{i,3} u_3(\lambda,\mu)  L^{y_3} + 
    \frac{1}{2} q_{i,3} (r_{i,3} u_3(\lambda,\mu)  L^{y_3})^2 \nonumber \\
       &+& r_{i,4} u_4(\lambda,\mu)  L^{y_4} + q(\beta_c,\lambda,\mu,L) \;
\end{eqnarray}
as our standard Ansatz. To simplify the notation, we identify $u_j=g_{1,j}$
here.
We set $r_{4,3}=1$ and $r_{3,4}=1$, where $i=3$ corresponds to the 
Binder cumulant $U_4$ and $i=4$ to $U_C$.  We have skipped terms 
that mix $u_3$ and $u_4$, since we simulated at parameters $(\lambda,\mu)$, 
where at least one of the scaling fields has a small amplitude.
Below, analyzing the data, we specify how we parametrize $u_3$ and $u_4$.
$q(\beta_c,\lambda,\mu,L)$ contains the corrections that decay with 
$L^{-\epsilon}$, where $\epsilon \gtrapprox 2$. 
For $Z_a/Z_p$, we assumed that there are only corrections due to the 
breaking of the symmetry by the cubic lattice.
We assume that the amplitude of this correction does not depend on 
$\mu$ and $\lambda$. As in our previous work, 
we take $\epsilon_2=2.023$ as numerical value of the exponent. 
In the case of the other three quantities
there are corrections with the exponent $\epsilon_1 = 2- \eta$ due to 
the analytic background of the magnetic susceptibility. 
We write the coefficient of these corrections
as linear functions of $\lambda$ and $\mu$.  
In the case of the Binder cumulant $U_4$ and the new cumulant, we 
do not take into account the correction due to the breaking of the symmetry 
by the lattice. We expect that 
it is at least partially taken into account by the term with the exponent 
$\epsilon_1$. For $\xi_{2nd}/L$, we expect
that  there is even a third correction, and that there are huge cancellations 
between the terms. Therefore 
we have added in this case a second correction. 
We took a constant amplitude and the exponent $\epsilon_2=2.023$.
Obviously, this Ansatz suffers from truncation errors. 
The effect of these errors
can be checked by varying the range of $\lambda$ and $\mu$ and the linear
lattice sizes $L$ that are taken into account.

\subsection{Two fixed points in close neighborhood}
\label{TwoFix}
Generically for a perturbation of a conformally invariant fixed 
point (see, for example, \cite{frad}) one gets
\begin{equation}
\label{RGeqXX}
\frac{\mbox{d} g_i}{\mbox{d} l} = y_i g_i - C_{kli} g_k g_l + ... \;,
\end{equation}
where $y_i$ is the RG exponent of the perturbation,
$C_{kli}$ is a structure constant, up to a constant factor, set by convention, 
and $g_i$ a dimensionless coupling.
Here we consider a single relevant perturbation with $0 < y \ll 1$. We get
\begin{equation}
\label{RGeq}
 \frac{\mbox{d} g}{\mbox{d} l} = y g - C g^2 + O(g^3)  \; .
\end{equation}
Note that the authors of Ref. \cite{AharonyNeu} discuss the same equation
[their Eq.~(11)], 
where $y$ and $C$ are obtained from the analysis of the six-loop 
$\epsilon$ expansion. See also Eq.~(27) of Ref. \cite{CB_O3}. 
Here $y$ and $C$ are free parameters that are 
eventually fixed by fitting numerical data.
We assume $g$ to be small and hence ignore the $O(g^3)$ contributions 
in the following. In addition to the fixed point $g=0$, 
there is the fixed point
\begin{equation}
\label{gstar}
g^* =  \frac{y}{C} \;.
\end{equation}
Let us rewrite Eq.~(\ref{RGeq}) by using $ \delta = g - g^*$:
\begin{equation}
 \frac{\mbox{d} \delta}{\mbox{d} l} =  y (g^* + \delta) - C (g^* + \delta)^2 
= - y \delta   - C \delta^2  \;.
\end{equation}
Hence at the fixed point $g=g^*$, there is an 
irrelevant perturbation with RG exponent $-y$.  

With respect to our finite size scaling study, we  vary the linear
lattice size $L$, while the coupling at the cutoff scale is given. 
The differential equation $g' = y g -C g^2$
is discussed in various contexts in the literature and one finds the 
solution 
\begin{equation}
g = \frac{y}{C + p \exp(- y l) } \;,
\end{equation}
where $p$ is an integration constant. Solving for
$p$, for given $g_0$ at the scale $\exp(l_0)$, we get
\begin{equation}
g = \frac{g^*}{1 + \left(\frac{g^*}{g_0} -1 \right) \exp(- y [l-l_0]) } \;.
\end{equation}
The coupling constant $g_0$ should be an analytic function of the parameters
of the model
\begin{equation}
\label{g0mu}
 g_0 = r(\lambda) \mu + s(\lambda) \mu^2 + ... \;.
\end{equation}
The dimensionless quantity $U_C$ at the critical point is an analytic 
function of $g$ at the scale proportional to $L=\exp(l)$, where $L$ is the 
linear size of the 
lattice
\begin{equation}
\label{UCg}
U_C(g) = A g + B g^2 + ... \; .
\end{equation}
Taking the leading order in Eqs.~(\ref{g0mu}) and (\ref{UCg}) only, 
we arrive at
\begin{equation}
\label{UCbasic}
U_C(\mu,\lambda,L) = 
 \frac{U_C^*}{1 + q \left(\frac{\mu^*}{\mu} -1 \right) L^{-y} } \;,
\end{equation}
where $U_C^*=A g^*$ and $r \mu^* = g^*$. The factor $q$ reflects the 
uncertainty of the identification of the length scales. $\mu^*$ and $q$ 
might depend on $\lambda$.

\section{The simulations and the analysis for $N=4$}
\label{O4simulations}
First we have simulated the model for $N=4$. Here the $O(N)$ invariant 
and the cubic fixed point are better separated than for 
$N=3$, which should make the analysis of the data more simple.
First we performed simulations for $\lambda=18.5$ at various
values of $\mu$.  Note that for the $O(4)$-invariant $\phi^4$ model 
$\lambda^*= 18.4(9)$  \cite{myON}. Extensive simulations for $\mu=0$,
were performed 
for $\lambda=18.5$ in Ref. \cite{myON}. In the preliminary stage of the
analysis we mainly monitored $\bar{U}_C$, which is  $U_C$ at  the value 
of $\beta$ such that $[Z_a/Z_p](\beta) =  0.11911$. 
Note that $(Z_a/Z_p)^*= 0.11911(2)$ for the $O(4)$-invariant $\phi^4$ model 
on a $L^3$ torus \cite{myON}.

The cubic fixed point is identified by $\bar{U}_C$ not depending on the 
linear lattice size $L$.  We arrived at $\bar{U}_C^* \approx -0.086$ and 
$\mu^* \approx 3.5$. 
However, a more detailed analysis of the data showed that for 
$(\lambda,\mu) = (18.5,3.5)$, corrections $\propto L^{-\omega_1}$ with 
$\omega_1 \approx 0.8$ have a considerable magnitude. 
Prompted by this result we started a more general search for 
$(\lambda,\mu)^*$, where both leading and subleading corrections are 
vanishing.  As preliminary estimate we arrived at 
$(\lambda,\mu)^* \approx (7,2.7)$.

In particular, to get accurate estimates of the correction exponents, 
we simulated at various values of $(\lambda,\mu)$, focussing on the 
neighborhood of $(\lambda,\mu)^*$. The data sets used in 
our final analysis, containing 20 different pairs $(\lambda,\mu)$,
are summarized in Table \ref{corrections_N4}.

For most of these pairs we simulated the linear lattice sizes 
$L=12$, $14$, $16$, $18$, $20$, $24$, $28$, $32$, $40$, $48$, and $56$. 
More, and in particular larger lattice sizes were added for 
$(\lambda,\mu)=(18.5,3.5)$,
$(18.5,4)$, $(7,2.64)$, and $(7,3)$  in order to determine the 
critical exponents $\nu$ and $\eta$.  In particular for 
$(\lambda,\mu)=(7,2.64)$, which is close to our final estimate of 
$(\lambda,\mu)^*$, we have simulated in addition $L=6$, $7$, $8$, $9$, 
$10$, $11$, $13$, $15$,  $64$, $72$, $80$, and 
$100$.
For example, for $(\lambda,\mu)=(7,2.64)$, we performed between
$10^9$ and $3  \times  10^9$ measurements for $L=6$ up to $32$. Then the
statistics is going down with increasing lattice size. For $L=100$, 
we performed $1.3 \times 10^8$ measurements.
In total we have used the equivalent of about $22$ years of CPU time
on a single core of an AMD EPYC$^{TM}$ 7351P CPU.

\subsection{Dimensionless quantities}
First we have analyzed the dimensionless quantities by using a 
joint fit of all four quantities that we have measured and all 20 
pairs of $(\lambda,\mu)$. To this end, we used the 
Ansatz~(\ref{RGansatz1}). We used $u_3(\lambda,\mu)$ and $u_4(\lambda,\mu)$
as free parameters for each pair $(\lambda,\mu)$.

\begin{figure}
\begin{center}
\includegraphics[width=14.5cm]{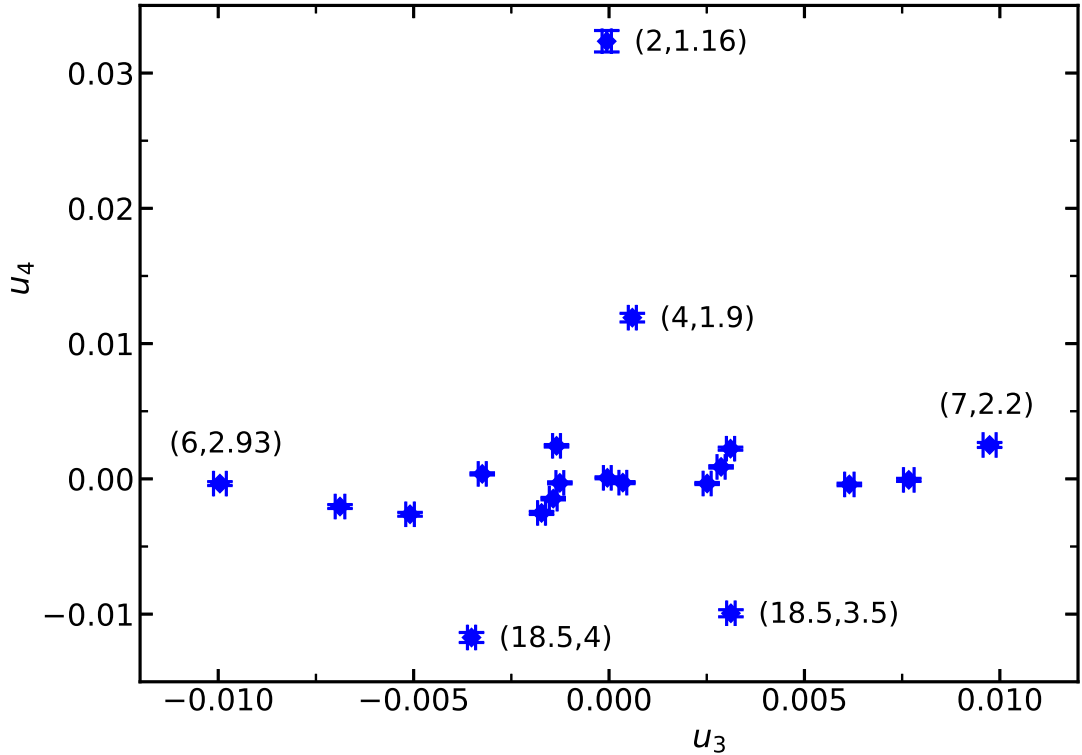}
\caption{\label{u3u4N4}
We plot the estimates of the correction amplitudes $(u_3,u_4)$ obtained by 
fitting our data for dimensionless quantities for $N=4$
and $L_{min}=12$ with the Ansatz~(\ref{RGansatz1}). Each data point 
corresponds to a pair $(\lambda,\mu)$ we simulated at.
To keep the figure readable, we give 
the values of $(\lambda,\mu)$ only for some data points. 
The complete information is given in Table \ref{corrections_N4}.
}
\end{center}
\end{figure}

Already for $L_{min}=12$ we find an acceptable $\chi^2/$DOF$=0.999$ 
corresponding to a $p$-value $p=0.504$.
In table \ref{corrections_N4} we give the correction amplitudes $u_3$ 
and $u_4$ for each $(\lambda,\mu)$, and the estimate 
of $\beta_c$. In the case of $u_3$ and $u_4$ we give the statistical error 
for this particular fit only. In Fig. \ref{u3u4N4} we plot the 
estimates of $u_3$ and $u_4$. 
Note that
we avoided values of $(\lambda,\mu)$, where both $u_3$ and $u_4$ have a large
modulus. This way we tried to reduce the effect of corrections that 
contain both scaling fields $u_3$ and $u_4$. 

In order to get the final estimate of $\beta_c$, the dimensionless quantities
$R_i$ and the correction exponents $\omega_1$ and $\omega_2$ and their error,
we produced a set of estimates with their respective statistical error.
To this end, we take  Ansatz~(\ref{RGansatz1}) for $L_{min}=12$ and $16$. 
Furthermore,
we extended Ansatz~(\ref{RGansatz1}) in four different ways or skipped one
pair of $(\lambda,\mu)$:
\begin{itemize}
\item
adding a term proportional to $L^{-4}$;
\item
adding a term proportional $u_3 L^{-\omega_2}$ to the third power;
\item
adding a term proportional $u_4 L^{-\omega_1}$ to the second power;
\item
adding a mixed $u_3 u_4 L^{-\omega_2 -\omega_1}$ term;
\item 
skipping $(\lambda,\mu)=(2,1.16)$ from the data.
\end{itemize}
These are all taken at $L_{min}=12$, giving all $\chi^2/$DOF $\approx 1$.
We compute the minimum $\beta_{c,min}$ of $\beta_c - error$ for each pair 
$(\lambda,\mu)$
among these different estimates. The same is done for the maximum 
$\beta_{c,max}$
of $\beta_c + error$. As our final estimate we take 
$(\beta_{c,max}+\beta_{c,min})/2$ and $(\beta_{c,max}-\beta_{c,min})/2$ as 
error. The results are given in the last column of Table 
\ref{corrections_N4}.

\begin{table}
\caption{\sl \label{corrections_N4}
Amplitudes $u_3$ and $u_4$ of the corrections obtained by fitting our data
for dimensionless quantities for $N=4$ 
for $L_{min}=12$ by using the Ansatz~(\ref{RGansatz1}).
In the last column, we give the estimate of 
$\beta_c$. Details are discussed in the text.
}
\begin{center}
\begin{tabular}{clrrc}
\hline
 \mc{1}{c}{$\lambda$} &  \mc{1}{c}{$\mu$} & \mc{1}{c}{$u_3$}  &  
\mc{1}{c}{$u_4$}  & \mc{1}{c}{$\beta_c$}  \\
\hline
  2   &  1.16 &--0.00006(12) &  0.03235(79) & 0.77776644(85) \\
  4   &  1.9  &  0.00060(10) &  0.01192(32) & 0.83415315(38) \\
  6   &  2.5  &--0.00134(10) &  0.00244(9)\phantom{0} & 0.85567074(29) \\
  6   &  2.93 &--0.00996(16) &--0.00035(15) & 0.84735549(45) \\
  6.5 &  2.4  &  0.00311(11) &  0.00223(11) & 0.86309955(34) \\
  6.5 &  2.7  &--0.00324(10) &  0.00037(8)\phantom{0}  & 0.85790473(32) \\
  7   &  2.2  &  0.00974(17) &  0.00251(18) & 0.87097872(53) \\
  7   &  2.5  &  0.00287(10) &  0.00087(9)\phantom{0}  & 0.86635289(24) \\
  7   &  2.64 &--0.00004(10) &  0.00008(7)\phantom{0}  & 0.86407506(17) \\
  7   &  2.7  &--0.00126(10) &--0.00030(8)\phantom{0}  & 0.86307673(28) \\
  7   &  3    &--0.00688(12) &--0.00205(15) & 0.85789401(27) \\
  7.2 &  2.656&  0.00035(10) &--0.00028(8)\phantom{0}  & 0.86566530(27) \\
  7.5 &  2.6  &  0.00251(10) &--0.00034(8)\phantom{0}  & 0.86913520(28) \\
  7.5 &  2.8  &--0.00143(10) &--0.00146(9)\phantom{0}  & 0.86598689(39) \\
  7.5 &  3    &--0.00509(11) &--0.00262(14) & 0.86270900(33) \\
  8   &  2.43 &  0.00767(14) &--0.00008(11) & 0.87542511(43) \\
  8   &  2.5  &  0.00615(12) &--0.00041(10) & 0.87443835(37) \\
  8   &  2.9  &--0.00173(10) &--0.00252(11) & 0.86849718(33) \\
 18.5 &  3.5  &  0.00312(11) &--0.00994(26) & 0.89931064(30) \\
 18.5 &  4    &--0.00352(10) &--0.01173(37) & 0.89496905(38) \\
\hline
\end{tabular}
\end{center}
\end{table}

In the case of the other quantities we proceed analogously.
We obtain
$\omega_2=0.082(5) $,  $\omega_1=0.763(24) $ for the correction exponents.
Note that the estimate of $\omega_1$ is within errors the same as 
$\omega=0.755(5)$ obtained for the $O(4)$-symmetric fixed point in 
Ref. \cite{myON}. The value of $\omega_2$ is clearly smaller than 
$Y_4= 0.125(5)$ obtained in \cite{O234}, indicating that the approximation
discussed in Sec. \ref{TwoFix} is not appropriate for $N=4$.
Our results for the dimensionless quantities are given in Table
\ref{dimensionless4}. These clearly differ from the $O(4)$-symmetric 
counterparts.

\begin{table}
\caption{\sl \label{dimensionless4}
Values of dimensionless quantities for a $L^3$ lattice with 
periodic boundary conditions for the cubic fixed point for $N=4$. 
For comparison we give the results obtained
in Ref. \cite{myON} for the $O(4)$ symmetric case.
}
\begin{center}
\begin{tabular}{clllc}
\hline
quantity &  \mc{1}{c}{$Z_a/Z_p$}  &   \mc{1}{c}{$\xi_{2nd}/L$}  &   \mc{1}{c}{$U_4$} &  $ U_C $ \\
\hline
 cubic, $N=4$ & 0.113495(41)&   0.56252(11)  & 1.104522(71) & -0.08869(22) \\
$O(4)$-symm.& 0.11911(2)  &   0.547296(26) & 1.094016(12) &   0   \\
\hline
\end{tabular}
\end{center}
\end{table}

Next we consider an Ansatz based on eq.~(\ref{RGansatz1}), 
where the scaling fields $u_3$ and $u_4$
are parameterized as quadratic functions of $\lambda$ and $\mu$
\begin{equation}
\label{para_u} 
u_j = a_{j,\lambda,1} (\lambda-\lambda^*) + a_{j,\mu,1} (\mu-\mu^*)
    + \frac{1}{2} a_{j,\lambda,2} (\lambda-\lambda^*)^2
    + \frac{1}{2} a_{j,\mu,2} (\mu-\mu^*)^2
    + a_{j,\lambda,\mu} (\lambda-\lambda^*) (\mu-\mu^*) \;.
\end{equation}
In our Ansatz, $\lambda^*$, $\mu^*$, and $a_{j,\lambda,1} $,
$a_{j,\mu,1}$, $a_{j,\lambda,2} $,  $a_{j,\mu,2}$, and
$a_{j,\lambda,\mu}$ for both values of $j$ are free parameters.

In order to get an acceptable $\chi^2/$DOF
we had to restrict the range of $\lambda$ and $\mu$ such that 7 or 8 
pairs $(\lambda,\mu)$ remained. 
Since this way data with a large amplitude of $u_3$ and $u_4$ are 
excluded, no accurate estimate of $\omega_1$ and $\omega_2$ is 
obtained in the fit. Therefore, we have fixed these to the values obtained 
above.

In order to get the final estimate we considered the following Ans\"atze 
and data sets:
Ansatz~(\ref{RGansatz1}) for $L_{min} =12$ and $16$ and the Ansatz~(\ref{RGansatz1}) with a term
proportional to $L^{-4}$ for $L_{min} =12$. Using these  Ans\"atze we fitted the data
set with  7 or 8 pairs $(\lambda,\mu)$. Based on these results, proceeding as discussed above, 
we arrive at
\begin{equation}
\label{parameterstar}
 (\lambda,\mu)^* = (7.10(15), 2.642(26)).
\end{equation}
Furthermore get $a_{3,\lambda,1}/a_{3,\mu,1}=-0.180(5)$, Eq.~(\ref{para_u}), 
characterizing the 
line of vanishing $u_3$ in the neighborhood of  $(\lambda,\mu)^*$.

Below we compute the exponents $y_t=1/\nu$ and $\eta$ based on our data for 
$(\lambda,\mu)=(7,2.64)$, 
which is close to $(\lambda,\mu)^*$. In order to estimate 
errors due to residual correction amplitudes $u_3$ and $u_4$, 
we compare with results obtained for $(\lambda,\mu)=(7,3)$ and
$(\lambda,\mu)=(18.5,3.5)$ and $(18.5,4)$, respectively.
Analyzing our estimates of the correction amplitudes obtained 
by using the different Ans\"atze discussed above, we find that
$|u_3|$ should be at least by a factor of 16 smaller for 
$(\lambda,\mu)=(7,2.64)$ than for
$(\lambda,\mu)=(7,3)$. $|u_4|$ should be at least by a factor of 16 smaller  
for $(\lambda,\mu)=(7,2.64)$ than for $(\lambda,\mu)=(18.5,3.5)$ and
$(18.5,4)$.

\subsection{The critical exponents $\eta$ and $\nu$}
Here we focus on the analysis of our data for $(\lambda,\mu)=(7,2.64)$, 
which is close to $(\lambda,\mu)^*$. In addition, we analyze
$(\lambda,\mu)=(7,3)$, $(18.5,3.5)$, and $(18.5,4)$
in order to estimate the possible effect of residual corrections at
$(\lambda,\mu)=(7,2.64)$.

\subsection{$\eta$ from the FSS behavior of the magnetic susceptibility}

\begin{figure}
\begin{center}
\includegraphics[width=14.5cm]{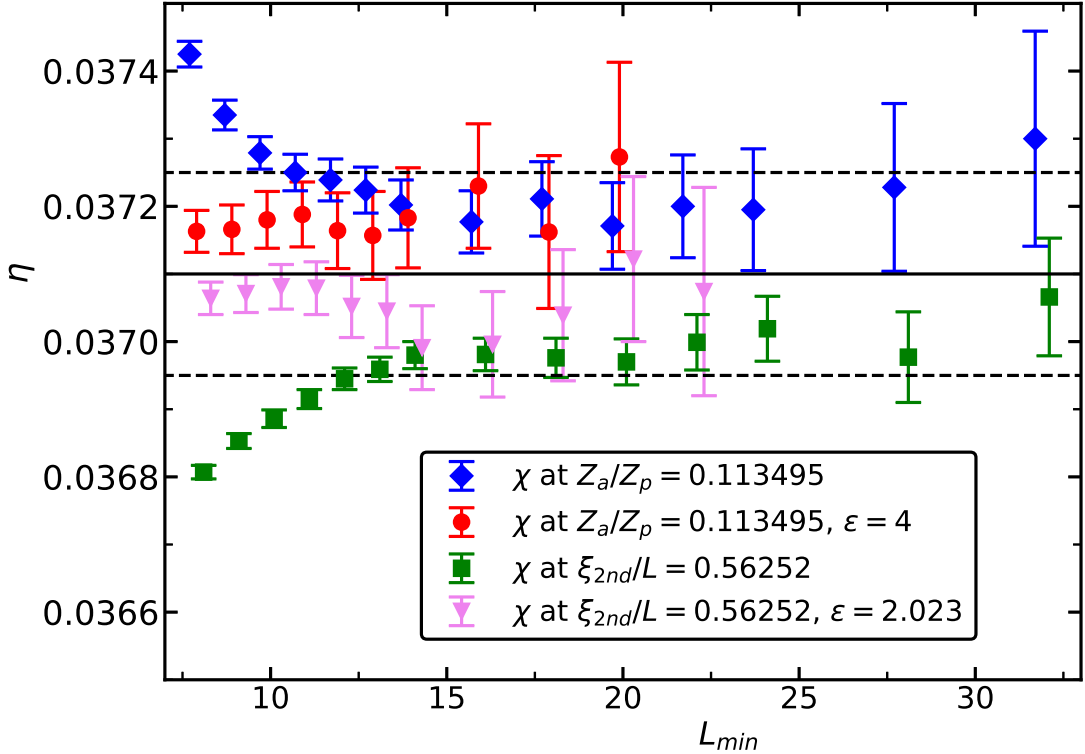}
\caption{\label{etaN4}
We plot the estimates of $\eta$ obtained by fitting the data for the 
magnetic susceptibility $\chi$ at
$(\lambda,\mu)=(7,2.64)$ and $N=4$ by using the 
Ans\"atze~(\ref{chifit1},\ref{chifit2}) versus the minimal lattice size
$L_{min}$ that is taken into account. 
The solid line gives our final estimate, while the dashed lines indicate our 
preliminary error estimate.
Note that the values on the $x$-axis are slightly shifted to reduce overlap
of the symbols.
}
\end{center}
\end{figure}

We have analyzed our data for the magnetic susceptibility at 
$(\lambda,\mu)=(7,2.64)$ 
at either $Z_a/Z_p=0.113495$ or $\xi_{2nd}/L=0.56252$.
We used the Ans\"atze
\begin{equation}
\label{chifit1}
 \bar{\chi} = a L^{2-\eta} + b
\end{equation}
or
\begin{equation}
\label{chifit2}
 \bar{\chi} = a L^{2-\eta} (1 + cL^{-\epsilon}) + b \;,
\end{equation}
where we have taken either $\epsilon=2.023$ or $4$.  Our results are plotted
in Fig. \ref{etaN4}. Our preliminary estimate $\eta=0.03710(15)$ is chosen such 
that all four fits are consistent with the estimate for some range of
$L_{min}$. 
In order to estimate the error due to residual correction amplitudes 
$u_3$ and $u_4$ at $(\lambda,\mu)=(7,2.64)$, 
we have analyzed the magnetic susceptibility at $(\lambda,\mu)=(7,3)$, 
$(18.5,3.5)$, and $(18.5,4)$ by using the same Ans\"atze as for 
$(\lambda,\mu)=(7,2.64)$.
In the case of $(\lambda,\mu)=(7,3)$, we see a larger spread between the 
results for $Z_a/Z_p=0.113495$ and $\xi_{2nd}/L=0.56252$ fixed. For
$\xi_{2nd}/L=0.56252$, using Ansatz~(\ref{chifit1}), we get very similar
results as for $(\lambda,\mu)=(7,2.64)$. In contrast for $Z_a/Z_p=0.113495$,
using Ansatz~(\ref{chifit1}), we get, for example, $\eta = 0.03753(8)$
for $L_{min}=16$. 

In the case of $(\lambda,\mu)=(18.5,3.5)$ and $(18.5,4)$,  
the estimates of $\eta$ are 
smaller than those for $(\lambda,\mu)=(7,2.64)$ throughout. 
For example, $\eta=0.03661(12)$
using Ansatz~(\ref{chifit1}) for $L_{min}=20$ and $(18.5,3.5)$, fixing 
$Z_a/Z_p=0.113495$. 

Given the discussion on the relative amplitude of the scaling fields
above, we enlarge the error to
\begin{equation}
\eta=0.0371(2) \;\;,
\end{equation}
to take into account the possible effect of residual corrections due 
to the scaling fields $u_3$ and $u_4$ at $(\lambda,\mu)=(7,2.64)$. Our 
estimate clearly differs from $\eta_{O(4)}= 0.03624(8)$, Ref. \cite{myON}
for the $O(4)$ symmetric fixed point.

\subsection{The thermal RG-exponent $y_t=1/\nu$ from the FSS behavior of the 
slopes of phenomenological couplings}

The slope of a dimensionless quantity at the critical point behaves as 
\begin{equation}
S_i = \frac{\partial R_i}{\partial \beta} = a_i L^{y_t} 
  (1 + \sum_j b_{i,j} L^{y_j} + ...) + \sum_j  c_{i,j} L^{y_j} + ... \;,
\end{equation}
where $b_{i,3}$ and $b_{i,4}$ vanish for an improved model, while 
$c_{i,3} $ and $c_{i,4}$ are finite. 

In order to check the effect of $c_{i,3} L^{y_3}$ 
we can construct linear combinations of dimensionless
quantities that do not depend on $u_3$. To this end we use the results
of the previous section, where we obtained the dependence of $R_i$ on 
$u_3$. In particular we have constructed such combinations for 
either $Z_a/Z_p$ or $\xi_{2nd}/L$ with $U_C$. 

\begin{figure}
\begin{center}
\includegraphics[width=14.5cm]{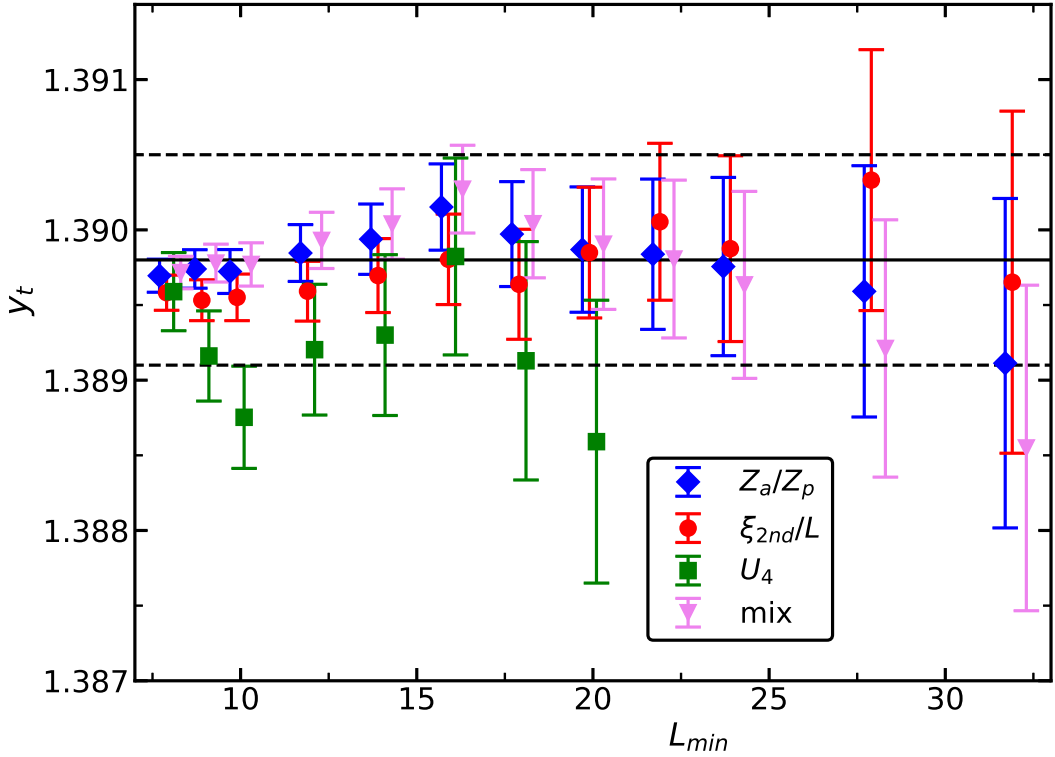}
\caption{\label{ytN4}
We plot the estimates of $y_t$ obtained by fitting the slopes of
$Z_a/Z_p$, $\xi_{2nd}/L$, $U_4$ and $Z_a/Z_p - 0.165 U_C$ at
$Z_a/Z_p =0.113495$
using the Ansatz~(\ref{slopeAnsatz}) with $\epsilon=2.023$
versus the minimal lattice size $L_{min}$ that is taken into account.
Data for $(\lambda,\mu)=(7,2.64)$ and $N=4$ are analyzed.
The dimensionless quantity is 
given in the legend, where ``mix'' refers to $Z_a/Z_p - 0.165 U_C$.
The solid line gives our final estimate, while the dashed lines indicate our
preliminary error estimate.
Note that the values on the $x$ axis are slightly shifted to reduce overlap
of the symbols.
}
\end{center}
\end{figure}

We have computed the slopes of dimensionless quantities at  
either $\xi_{2nd}/L=0.56252$ or $Z_a/Z_p =0.113495$.
We have fitted our data with the Ansatz
\begin{equation}
\label{slopeAnsatz}
 \bar{S} = a L^{y_t}  (1 + c L^{-\epsilon}) \;,
\end{equation}
where we take $\epsilon=2.023$, which is the estimate of the exponent
related with the violation of the rotational invariance by the lattice.
As a check, we performed fits with $\epsilon=2-\eta$, 
taking our estimate of $\eta$
obtained above. The estimates of $y_t$ change only by little.  
In Fig. \ref{ytN4} we plot the estimates for $Z_a/Z_p =0.113495$
obtained by fitting the 
data for $(\lambda,\mu)=(7,2.64)$ taking  $\epsilon=2.023$. 
As preliminary result we obtain $y_t = 1.3898(7)$. It is chosen such that 
the estimates obtained by the fits are covered for all four slopes 
for some range of $L_{min}$. Analyzing the slopes at $\xi_{2nd}/L=0.56252$,
we get fully consistent results.

We have repeated this analysis for $(\lambda,\mu)=(7,3)$, $(18.5,3.5)$, and
$(18.5,4)$ to see the effect of the corrections on the estimate of $y_t$.
For $(\lambda,\mu)=(7,3)$ we get essentially consistent results from the 
different slopes that we consider. We get the estimate 
$y_t \approx  1.396$ 
being clearly larger than the estimate obtained for $(\lambda,\mu)=(7,2.64)$.

In the case of $(18.5,3.5)$ and $(18.5,4)$ the estimate of $y_t$ obtained 
from the slope of $U_4$ is clearly larger than that obtained from 
the slopes of $Z_a/Z_p$ and $\xi_{2nd}/L$.  Likely this is due to the 
fact that the effect of a finite scaling field $u_4$ is different in the 
different slopes. On top of this, there is a clear difference between 
the results of $(18.5,3.5)$ and $(18.5,4)$, which we attribute to the 
different sign of $u_3$ for these two values of $(\lambda,\mu)$. 
From the slopes of $Z_a/Z_p$ and $\xi_{2nd}/L$  we get 
$y_t \approx 1.386$ and $1.392$, respectively.

Given the discussion on the relative amplitude of the scaling fields
above, we enlarge the error to
\begin{equation}
y_t = 1.3898(13)
\end{equation}
to take into account the possible effect of residual corrections due
to the scaling fields $u_3$ and $u_4$ at $(\lambda,\mu)=(7,2.64)$.

\section{Critical exponents $Y_l$ for $O(3)$ symmetry}
\label{O3exponents}
We have extended the simulations of Ref. \cite{O234} focussing on $N=3$. 
We make use of the  estimate $\lambda^*=5.17(11)$ given in the Appendix of
\cite{myIco}. Close to $\lambda^*$, the inverse critical temperature is 
estimated as $\beta_c(\lambda=5.0) = 0.68756127(13)[6]$ and 
$\beta_c(\lambda=5.2) = 0.68798521(8)[3]$. 

Here we study the same quantities as in Ref. \cite{O234}. 
We consider perturbations $P_{m,l}$ defined by the power $m$ of
the order parameter and the spin representation $l$ of the O($N$) group
\begin{equation}
  P_{m,l}^{a_1...a_l}(\vec{\Phi}) = (\vec{\Phi}^{\, 2})^{(m-l)/2} Q_l^{a_1...a_l}(\vec{\Phi}) \;,
\label{pmldef}
\end{equation}
where $Q_l^{a_1...a_l}$ is a homogeneous polynomial of degree $l$
that is symmetric and traceless in the $l$ indices. For $l=4$, see 
Eq.~(\ref{spin4}) above.
We consider correlators of the type  
\begin{equation}
C_l = \sum_{a_1,a_2,...,a_l}  
\left \langle \sum_x Q_l^{a_1,a_2,...,a_l} (\vec{\phi}_x)   \:
Q_l^{a_1,a_2,...,a_l} (\vec{m}) \right \rangle \;,
\end{equation}
where $\vec{M}=\sum_x \vec{\phi}_x$ and $\vec{m}=\vec{M}/|\vec{M}|$. 
And in addition
\begin{equation}
D_l = \frac{ \sum_{a_1,a_2,...,a_l}  \left \langle \sum_x Q_l^{a_1,a_2,...,a_n} (\vec{\phi}_x)   \:
Q_l^{a_1,a_2,...,a_l} (\vec{M}) \right \rangle} 
{ \langle \vec{M}^{\, 2} \rangle^{l/2} } \;.
\end{equation}

In terms of the angle $\alpha$ between $\vec{m}$ and $\vec{\phi}_x$ defined by
\begin{equation} 
\vec{\phi}_x \cdot \vec{m} = |\vec{\phi}_x| \cos(\alpha_x) 
\end{equation}
one gets, for example
\begin{equation}
C_4 = \left \langle \sum_x |\vec{\phi}_x|^4 \left(\cos^4 \alpha_x -\frac{6}{N+4} \cos^2 \alpha_x 
  + \frac{3}{(N+2) (N+4)}   \right) \right \rangle \;.
\end{equation}
The new simulations were in particular designed for $l=4$.  Furthermore
we have added measurements for $l=5$ and $6$.  
We notice that the estimators $C_l$ and $D_l$
become increasingly noisy with increasing $l$. 
This means that integrated autocorrelation times $\tau_{int}$ go to $0.5$,
while the relative variance increases as the lattice size $L$ increases.
This behavior can be seen starting from $l=4$. 
Here we try to attenuate the problem by frequent measurements. To 
this end, we have implemented local updates, in particular the 
over-relaxation update efficiently by using AVX intrinsics. 
See Sec. \ref{algorithms}.

The most recent update and measurement cycle is
\begin{verbatim}
rotate();
for(i=0;i<N_cl;i++) {cluster(0); cluster(1); cluster(2);}
metro(); measure_ene(); measure_X();
for(i=0;i<N_ov;i++) {over(); measure_X();}
\end{verbatim}
\verb+rotate()+ is a global rotation of the field $\phi$ by a random 
$O(3)$ matrix.
\verb+cluster(i)+ is a single-cluster update of the $i^{th}$ component 
of the field.
\verb+metro()+ is the local Metropolis update sweeping over the lattice.
At each site an over-relaxation update follows the Metropolis update as second hit.
\verb+over()+ is a sweep with the over-relaxation
update. \verb+measure_ene()+ is the measurement of the energy, Eq.~(\ref{energy}). 
It remains unchanged under over-relaxation updates.
\verb+measure_X()+ is the measurement of the magnetic susceptibility 
[Eq.~(\ref{suscept})] $C_l$ and $D_l$. 
In the most recent simulations, 
we used  \verb+N_ov+ $=20$ and \verb+N_cl+$=L/8$. Some of the 
simulations for $L < 30$ were performed without cluster updates.

We performed simulations at $\lambda=5.2$ and $\beta=0.68798521$ for the linear lattice sizes 
$L=6$, $7$, ..., $28$, $30$, $32$, $36$, $40$ and $48$. In Ref. \cite{O234}
larger lattice sizes have been simulated. However to get an accurate estimate
of $Y_4$ it is better to generate high statistics for relatively small
lattice sizes.

In the case of our largest linear lattice size $L=48$ we performed 
$410\;320\;000$ cycles for four copies of the field, while 
for linear lattice sizes up to $L \approx 20$  about $2 \times 10^9$
cycles for four copies of the field are performed.  Going from
$L \approx 20$ up to $L=48$ the statistics gradually decreases.

To check the effect of $\lambda$ on our numerical result we performed 
simulations at $\lambda=5$ and $\beta=0.68756127$ for linear lattice sizes
up to $L=24$. 

In total we have used about the equivalent of 
20 years of CPU time
on a single core of an AMD EPYC$^{TM}$ 7351P CPU.

\subsection{Analysis of the data}

\begin{figure}
\begin{center}
\includegraphics[width=14.5cm]{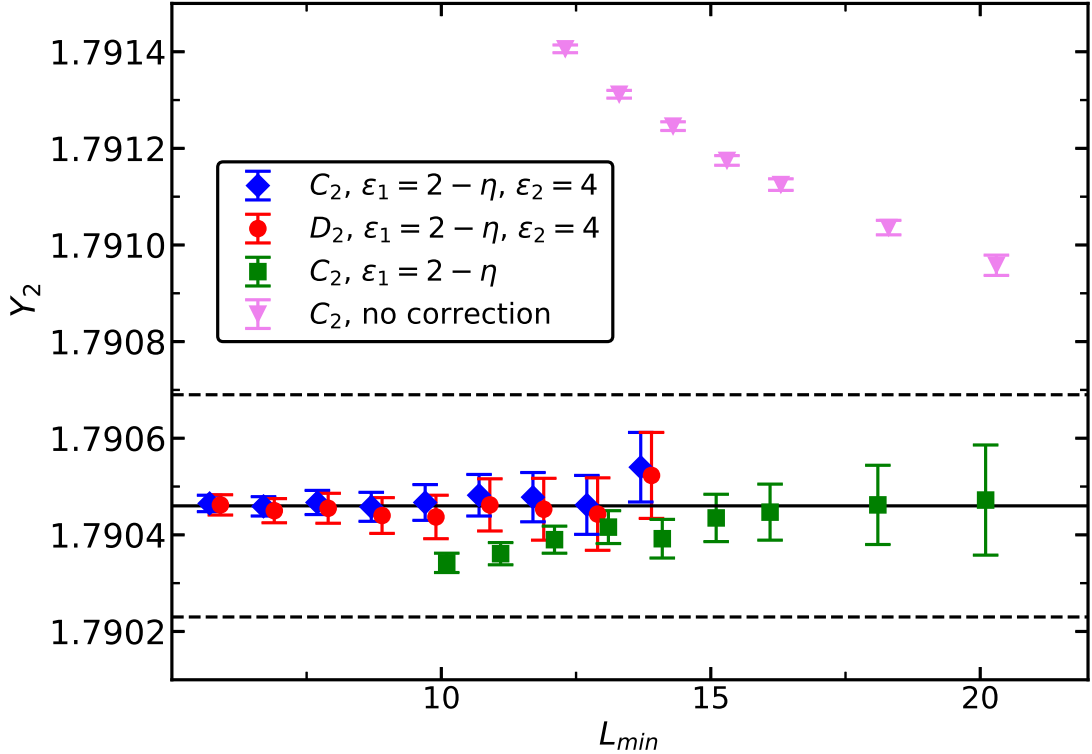}
\caption{\label{Y2}
We plot the estimates of $Y_2$ for $N=3$ obtained by fitting our data
for $C_2$ and $D_2$ at $\lambda=5.2$ by using the 
Ans\"atze~(\ref{nocorr},\ref{corr2},\ref{corr4}).
Note that the values on the $x$-axis are slightly shifted to reduce overlap
of the symbols. The solid line gives the estimate of Ref. \cite{CB_O3},
while the dashed lines indicate the error.
}
\end{center}
\end{figure}

In Fig. \ref{Y2}  we show our estimates obtained for $Y_2$ by using the 
Ans\"atze
\begin{equation}
\label{nocorr}
C_l = a L^{Y_l} \;,
\end{equation}
\begin{equation}
\label{corr2}
C_l = a L^{Y_l} \left( 1  + b L^{-2+\eta} \right) 
\end{equation}
and
\begin{equation}
\label{corr4}
C_l = a L^{Y_l} \left( 1  + b L^{-2+\eta} + c L^{-4} \right) \;,
\end{equation}
where the term $c L^{-4}$ is an ad hoc choice that is justified by the 
improved quality of the fits and the fact that for $l=2$, the estimates
obtained by using the Ans\"atze~(\ref{corr2},\ref{corr4})
are in nice agreement with the result obtained by using the 
CB method \cite{CB_O3}. Analogous fits are performed for $D_l$. 
In general, the results obtained by fitting $D_l$ and $C_l$ are 
consistent. The statistical error is slightly smaller for $C_l$. 
Based on fits with the Ans\"atze~(\ref{corr2}) and (\ref{corr4}), we take
$Y_2=1.79047(11)$ as preliminary estimate.

As a check we reanalyzed our data for $\lambda=5.2$ at 
$\beta=0.68798521 \pm 0.00000011$, which is our estimate of $\beta_c \pm$ the 
estimate of the error.  We find that the estimate of $Y_2$ changes by 
about $2 \times 10^{-5}$ with some dependence on the type of the fit and 
on $L_{min}$. 
Furthermore, we have replaced $\epsilon_1=2-\eta$ by $\epsilon_1=2.023$.
Also here, the results change by about $2 \times 10^{-5}$.

Finally, we estimate the effect of residual leading-order
corrections to scaling due to the fact that $\lambda=5.2$ is only
an approximation of $\lambda^*$.
To this end we have simulated the linear lattice sizes $L=10$, $12$, ...,
$24$ at $\lambda=5.0$ and the estimate of $\beta_c$, $\beta=0.68756127$.
We computed the ratios $r_l(L)=C_l(L,\lambda=5.0)/C_l(L,\lambda=5.2)$.
We have fitted these ratios by using the Ansatz
\begin{equation}
r_l(L)=c L^{x} \;.
\end{equation}
Taking into account all lattice sizes that we simulated for $\lambda=5.0$
we get $x=0.000138(22)$ for $l=2$. 
We assume that the difference in the numerical estimate in the
exponent is dominated by the difference in the leading correction.
Based on the estimate $\lambda^*=5.17(11)$, we
assume as lower bound $\lambda^* \ge 5.06$ in our estimate of the error.

Taking these different errors into account we arrive at the final estimate 
\begin{equation}
Y_2=1.7905(3) \;.
\end{equation}

Performing a similar analysis, we arrive at 
\begin{equation}
Y_3 = 0.9615(3) \;.
\end{equation}

\begin{figure}
\begin{center}
\includegraphics[width=14.5cm]{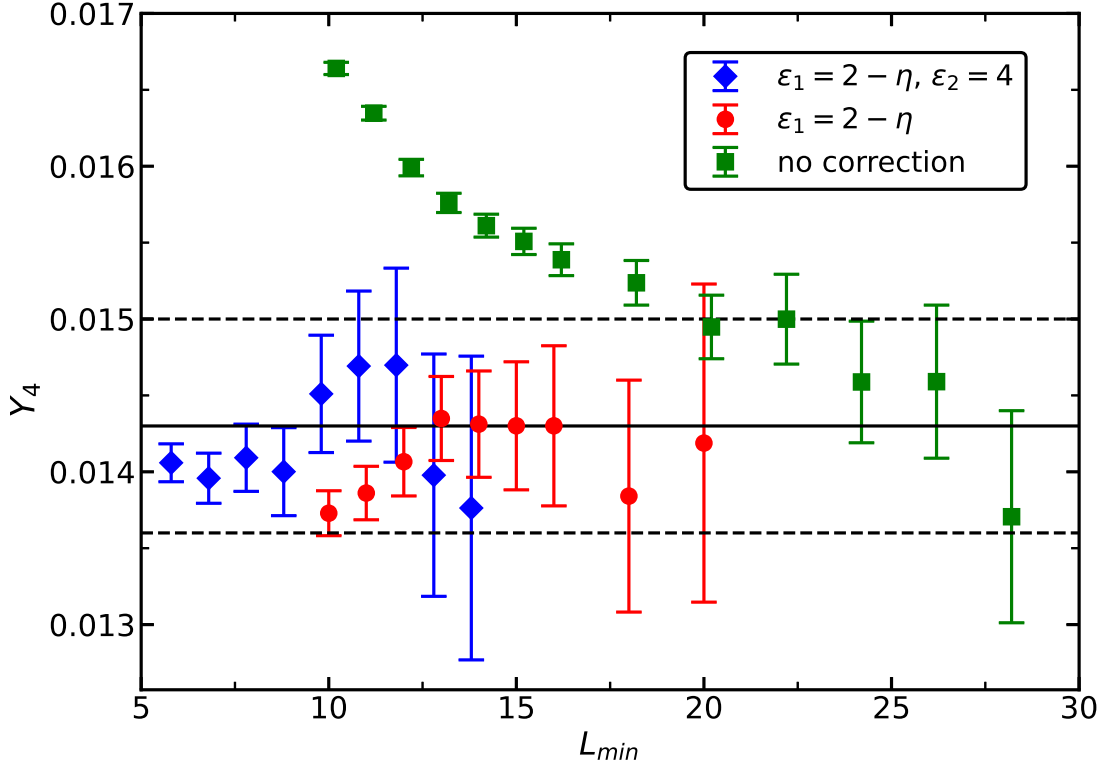}
\caption{\label{C4}
We plot the estimates of $Y_4$ obtained by fitting our data for $C_4$ 
at $\lambda=5.2$ for $N=3$ by using the 
Ans\"atze~(\ref{nocorr},\ref{corr2},\ref{corr4}).
Note that the values on the $x$-axis are slightly shifted to reduce overlap
of the symbols.
The solid line gives our preliminary estimate of $Y_4$,
while the dashed lines indicate the error.
}
\end{center}
\end{figure}

In Fig.~\ref{C4}   we plot results obtained for $Y_4$.
As preliminary estimate we take $Y_4=0.0143(7)$. Taking into account systematic
errors as discussed above for $l=2$, we arrive at the final estimate
\begin{equation}
Y_4 = 0.0143(9)  \;.
\end{equation}
In a similar fashion we arrive at $Y_5=-1.04(1)$ and $Y_6=-2.2(2)$. 
We notice that the error rapidly increases with increasing 
$l$. For $l \ge 5$ alternative approaches are likely more suitable.
See for example Refs. \cite{DebasishO2,DebasishO4}.

In these simulations we also have computed the magnetic susceptibility $\chi$.
Just as a check, we have fitted the data for $\lambda=5.2$ with the Ansatz
\begin{equation}
\chi = a L^{2-\eta} + b \;.
\end{equation}
We get an acceptable goodness of the fit starting from $L_{min}=8$.  
Estimates of the exponent are for example $\eta=0.037935(15)$, 
$0.037890(20)$, $0.037881(26)$, $0.037894(35)$, $0.037864(45)$, for 
$L_{min}=8$, $10$, $12$, $14$, and $16$. Starting from $L_{min}=10$ 
these estimates  are consistent 
with $\eta=0.037884(102)$ \cite{CB_O3} and $\eta=0.03784(5)$  \cite{myIco}.

\section{Simulations and analysis of the data for $N=3$}
\label{N3simulations}
We simulated at values of $\lambda$ that are close to
$\lambda^* = 5.17(11)$ of the $O(3)$-symmetric case $\mu=0$ \cite{myIco}.
In particular, we simulated the model at $\lambda=5.2$,
$\mu=-0.5$, $-0.3$, $-0.1$, $0.05$, $0.1$, $0.2$, $0.3$, 
$0.4$, $0.5$, and $0.7$ using $L_{max}=48$, $48$, $48$, $48$, $48$, 
$48$, $200$, $48$, $48$, $48$, respectively. Furthermore, we simulated
at $\lambda=5$, $\mu=-0.3$, $-0.1$, $0.1$, $0.2$, $0.25$,
$0.3$, and $0.4$,  using $L_{max}=96$, $48$, $48$, $48$, $100$, $200$, and
$48$, respectively,
at $\lambda =4.8$,  $\mu=-0.5$, $-0.3$, $0.2$, $0.3$, $0.4$,
and $0.5$, using $L_{max}=64$, $48$, $48$, $200$, $48$, and $64$, respectively,
at $\lambda =4.7$,  $\mu=-0.7$, and $0.7$ using $L_{max}=64$,
at 
$\lambda=4.5$, $\mu=-1.0$, $-0.5$, $-0.3$, $-0.15$, $0.15$, $0.25$, $0.3$, 
$0.35$, $0.5$ and $1.0$, using $L_{max}=64$, $32$, $32$, $32$, $32$, $32$, 
$32$, $32$, $32$, and $64$, respectively, and
at $\lambda =4.3$, $\mu=-1.0$ using $L_{max}=64$. 
Here, $L_{max}$ is the largest linear lattice size that we simulated.

For example for $(\lambda,\mu)=(5.0,0.3)$, which is close to $(\lambda,\mu)^*$
as we shall see below, we simulated the linear lattice sizes 
$L=8$, $9$,  ..., $16$, $18$, ..., $36$, $40$, ..., $48$, $56$, $60$, $64$, 
$72$, $80$, $90$, $100$, $110$, $120$, $140$, and $200$. We performed about
$2 \times 10^9$ measurements for each lattice size up to $L=24$. Then the 
number of measurements gradually drops. For example, we performed 
$8.6 \times 10^8$, $1.2 \times 10^8$, and $2.1 \times 10^7$ measurements
for $L=48$, $100$, and $200$, respectively. The simulations at 
$(\lambda,\mu)=(5.0,0.3)$ took about   
the equivalent of 25 years of CPU time
on a single core of an AMD EPYC$^{TM}$ 7351P CPU. All simulations for 
$\mu \ne 0$ took about the equivalent of 130 years of CPU time
on a single core of an AMD EPYC$^{TM}$ 7351P CPU. 

In addition we used the data for $\mu=0$,  $\lambda=5.0$ and $5.2$, 
discussed in the Appendix A of Ref. \cite{myIco}. We have added the lattice
sizes $L=32$, $48$, $56$, $64$, and $72$ for $\lambda=5.0$ and 
$L=32$, $48$, $56$, $64$, $72$ and $90$ for $\lambda=5.2$ to have a better
match with the lattice sizes simulated for $\mu \ne 0$.

Note that here we have simulated also negative values of $\mu$, where 
a first order transition is expected. However for the values of $\mu$ studied
here, the linear lattice size $L$ should be smaller by several orders than 
the correlation length at the transition. Therefore, it is justified to treat
the systems as if they were critical.

\subsection{dimensionless quantities} 
As for $N=4$, we first analyze the behavior of dimensionless quantities.  
In a first series of fits, we use Ans\"atze, where we expand around the 
$O(3)$-symmetric fixed point. For the dimensionless quantities $Z_a/Z_p$, 
$\xi_{2nd}/L$ and $U_4$ we take 
\begin{equation}
\label{R3master}
R_i(\beta_c,\lambda,\mu,L) =R_i^* +  r_{i,4}  w_4(\lambda,\mu) L^{-\omega} 
+ \sum_{m=2}^{m_{max}} c_{i,m} U_C^m(\beta_c,\lambda,\mu,L)
                           + \sum_j a_{i,j} L^{-\epsilon_j} \;,
\end{equation}
where $\epsilon_j \gtrapprox 2$. Equation~(\ref{R3master}) is a standard 
Ansatz for
analyzing dimensionless quantities at $\mu=0$, augmented by
$\sum_{m=2}^{m_{max}} c_{i,m} U_C^m(\beta_c,\lambda,\mu,L)$.
The basic idea of the Ansatz is that  
$Z_a/Z_p$, $\xi_{2nd}/L$ and $U_4$ behave as $R_i(\beta_c,\lambda,\mu,L)
=R_i(\beta_c,\lambda,0,L) + O(\mu^2)$, while $U_C=O(\mu)$, due to symmetry. 
Hence $R_i^*$ are the $O(3)$-symmetric 
fixed point values. Here we avoid an explicit parameterization 
of the RG flow of the cubic perturbation. Instead, we take it from the 
dimensionless quantity $U_C$. In our fits, we chose either $m_{max}=3$, $4$,
or $5$. 
The term $r_{i,4}  w_4(\lambda,\mu) L^{-\omega}$ is an approximation 
based on the fact that $\omega_1 \gg Y_4, \omega_2$. 
In the approximation 
we assume a line of fixed points, and furthermore that the correction exponent
$\omega_1$ stays constant along this line.

In order to fix the normalization of $w_4(\lambda,\mu)$, 
we set $r_{U_4,4}=1$ for the Binder cumulant $U_4$. The choice of 
subleading corrections depends on the dimensionless quantity. In the case
of $Z_a/Z_p$ we take $\epsilon=2.023$, which is an estimate of the correction
exponent related with the violation of rotational invariance by the lattice. 
The amplitude $a_{Z_a/Z_p,1}$ is assumed to be constant in $\lambda$ and $\mu$.
In the case of $\xi_{2nd}/L$, we take two correction terms, 
one with the correction exponent
$\epsilon_1=2-\eta$, associated with the analytic background of the magnetic 
susceptibility, and, as for $Z_a/Z_p$, one with the correction exponent 
$\epsilon_2=2.023$. The 
amplitude $a_{\xi_{2nd}/L,2}$ is assumed to be constant in $\lambda$ and $\mu$. 
The amplitude $a_{\xi_{2nd}/L,1}$ is parameterized as linear in 
$\lambda$ and quadratic in $\mu$. 
We experimented with various dependencies on $\lambda$ and $\mu$, which however
did not improve the quality of the fit. In the case of $U_4$, we take one 
correction term with $\epsilon_1=2-\eta$. 
The amplitude $a_{U_4,1}$ is parameterized as $a_{\xi_{2nd}/L,1}$. As check, 
we added a second correction term with $\epsilon_2=2.023$ in some of the fits.
We performed fits fixing 
$\omega=0.759$, which is the value obtained for the $O(3)$-symmetric 
fixed point \cite{myIco}. 
For technical reasons, we ignore the statistical error of 
$U_C(\beta_s,\lambda,\mu,L)$
and the Taylor coefficients in $(\beta_c-\beta_s)$. This is justified by the
fact that $\sum_{m=2}^{m_{max}} c_{i,m} U_C(\beta_c,\lambda,\mu,L)^m$
assumes only rather small values. As check, we
added a term proportional to $L^{-4}$ for each dimensionless quantity,
were the amplitudes are constant in $\mu$ and $\lambda$. 
In a first series of fits, we used $w_4(\lambda,\mu)$ as a 
free parameter for each pair $(\lambda,\mu)$.

Fitting the data for $|\mu| \le 0.5$ with the Ansatz~(\ref{R3master}) and 
$m_{max}=3$, we get $\chi^2/$DOF$=1.125$,  $1.076$, $1.086$ and
$1.066$ corresponding to $p=0.002$, $0.056$, $0.057$, and $0.13$
for $L_{min}=12$, $16$, $20$, and $24$, respectively.  
Adding 
a term proportional to $L^{-4}$ for each dimensionless quantity, we get 
$\chi^2/$DOF $=1.061$, $1.074$, $1.075$, and $1.061$ corresponding to 
$p=0.078$, $0.062$, $0.084$, and $0.148$ for $L_{min}=12$, $16$, $20$, and $24$,
respectively.
Fitting the data for $|\mu| \le 0.7$ with the Ansatz~(\ref{R3master}) and 
$m_{max}=3$, the $p$-value is smaller than $0.1$ for $L_{min}<24$, while 
for $L_{min}=24$ we get $\chi^2/$DOF $= 1.071$ corresponding to $p=0.106$. 
Adding 
a term proportional to $L^{-4}$ for each dimensionless quantity the fits
for $L_{min}<24$ are worse than those for  $|\mu| \le 0.5$, while for 
$L_{min}=24$ we get $\chi^2/$DOF $= 1.068$ corresponding to $p=0.116$. 

Fitting the data for $|\mu| \le 1.0$ with the Ansatz~(\ref{R3master}) and
$m_{max}=3$, for $L_{min}=24$ we get $\chi^2/$DOF $= 1.471$ corresponding 
to $p=0.000$.  Adding 
a term proportional to $L^{-4}$ for each dimensionless quantity, the quality of
the fit does not improve considerably.

Fitting the data for $|\mu| \le 1.0$ with the Ansatz~(\ref{R3master}) and
$m_{max}=4$, the quality of the fit improves considerably. We get
$\chi^2/$DOF $=1.111$, $1.065$, $1.078$, and $1.051$, corresponding to
$p=0.004$, $0.077$, $0.066$, and $0.179$ for $L_{min}=12$, $16$, $20$, and $24$,
respectively.  Adding
a term proportional to $L^{-4}$ for each dimensionless quantity, we 
get $\chi^2/$DOF $=1.050$, $1.063$, $1.070$, and $1.047$, corresponding to
$p=0.112$, $0.082$, $0.087$, and $0.198$ for 
$L_{min}=12$, $16$, $20$, and $24$, respectively.  
Going from $m_{max}=4$ to $m_{max}=5$, taking into account the data with 
$|\mu| \le 1.0$, the quality of the fits only slightly improves.

We conclude that our approximative Ansatz~(\ref{R3master}), for our high
statistics data, is at the edge of being acceptable, which in the literature
is usually assumed to be the case for  $0.1 \le p \le 0.9$. 
For $|\mu| \le 0.5$, $m_{max}=3$ seems to be sufficient, while for 
$|\mu| \le 1$ at least one more power of $U_C$ has to be added.

In Table \ref{O3fix} we give a few characteristic results for the 
dimensionless quantities
obtained by using these fits. 
These  are consistent with those of Ref. \cite{myIco}. A more accurate
final result than that given in Ref. \cite{myIco} can not be obtained.

\begin{table}
\caption{\sl \label{O3fix}
Estimates of the fixed point values $R^*$ of dimensionless 
quantities $R$ at the $O(3)$-invariant fixed point.
These are obtained by using the Ansatz~(\ref{R3master}). In the last line 
we give the final results of Ref. \cite{myIco} for comparison. For a 
discussion see the text.
}
\begin{center}
\begin{tabular}{cccclll}
\hline
$m_{max}$ & $L^{-4}$ & range &$L_{min}$ & \mc{1}{c}{$(Z_a/Z_p)^*$} & 
\mc{1}{c}{$(\xi_{2nd}/L)^*$ } & \mc{1}{c}{$U_4^*$} \\
\hline
3  & no  & $|\mu| \le 0.5$  & 24 & 0.194766(13)&0.564036(11) & 1.139284(10)\\
3  & yes & $|\mu| \le 0.5$  & 12 & 0.194753(7) &0.564051(6)  & 1.139299(6) \\
4  & no  & $|\mu| \le 1.0$  & 24 & 0.194761(11)&0.564041(10) & 1.139289(9) \\
4  & yes & $|\mu| \le 1.0$  & 12 & 0.194750(6) &0.564053(5)  & 1.139300(5) \\
\hline
   \mc{4}{c}{ref. \cite{myIco} }   & 0.19477(2)  & 0.56404(2)  & 1.13929(2)  \\
\hline
\end{tabular}
\end{center}
\end{table}

Next we consider the coefficients $c_{i,m}$.  As final result we quote 
numbers that are consistent with four different fits. First we took 
$m_{max}=4$ and data for $|\mu| \le 0.7$. Using a correction term proportional
to $L^{-4}$ we took the result for $L_{min}=12$, while without this correction
the data are taken for $L_{min}=24$. The third and fourth fit are analogous, 
but for $m_{max}=5$ and data for $|\mu| \le 1.0$.
We get $c_{Z_a/Z_p,2}=-0.64(5)$,  $c_{Z_a/Z_p,3}=2.1(3)$,
$c_{\xi_{2nd}/L,2}=1.34(4)$,  $c_{\xi_{2nd}/L,3}=-3.4(3)$,
$c_{U_4,2}=1.25(3)$, and $c_{U_4,3}=-3.0(3)$. Coefficients for $m=4$ and $5$
have large error bars and vary considerably among the different fits.

In the approximation used in the fits discussed in this section, there 
is an improved line $\lambda^*(\mu)$, where the correction proportional 
to $L^{-\omega}$ vanishes. We have computed zeros of $w_4(\lambda,\mu)$ for
given $\mu$ by linear interpolation in $\lambda$.  Our final results, which are 
consistent with the four different fits used above are given 
in table \ref{N3ls}. The maximum of $\lambda^*(\mu)$ is reached for $\mu=0$.
The result for  $\lambda^*(0)$ is consistent with $\lambda^*=5.17(11)$
obtained in ref. \cite{myIco}. $\lambda^*(\mu)$ is almost even in $\mu$. 
$\lambda^*(\mu)$ for negative values of $\mu$ is slightly smaller than for
the corresponding positive values of $\mu$.

\begin{table}
\caption{\sl \label{N3ls}
Numerical results for $\lambda^*(\mu)$ for $N=3$.
For a discussion see the text.
}
\begin{center}
\begin{tabular}{rl}
\hline
 \mc{1}{c}{$\mu$} &  \mc{1}{c}{$\lambda^*$} \\
\hline
  -0.7 &  4.53(22) \\
  -0.5 &  4.78(13) \\
  -0.3 &  4.97(10)  \\
  -0.1 &  5.08(10)  \\         
   0.0 &  5.10(10)  \\
   0.1 &  5.09(10)  \\           
   0.2 &  5.04(10)  \\
   0.3 &  4.98(10) \\        
   0.4 &  4.90(10) \\
   0.5 &  4.81(13) \\     
   0.7 &  4.55(22) \\   
\hline
\end{tabular}
\end{center}
\end{table}

Next we used the parameterization for the correction amplitude
\begin{equation}
\label{favoritew4}
 w_4(\lambda,\mu) = a (\lambda - \lambda^* - c \mu^2 -d  \mu^3) \;
  (1+e (\lambda-5.0)) 
\end{equation}
and
\begin{equation}
\label{favoritew4x}
 w_4(\lambda,\mu) = a (\lambda - \lambda^* - c \mu^2 -d  \mu^3 -e \mu^4) \;
  (1+f (\lambda-5.0) ) \;,
\end{equation}
where we have added one term proportional to $\mu^4$.  We obtain a similar
quality of the fit as above without parameterization of $w_4(\lambda,\mu)$.
Also the differences between
the two parameterizations~(\ref{favoritew4}, \ref{favoritew4x}) is minor.
Therefore we abstain from a detailed discussion. Let us just briefly summarize
the results for the parameters of Eqs.~(\ref{favoritew4}) and
(\ref{favoritew4x})
and the estimate of $\beta_c$ that we obtain.

To this end let us discuss the results of four selected fits
\begin{itemize}
\item
For the Ansatz~(\ref{R3master}) without a correction of $U_4$
proportional to $L^{-2.023}$ and no correction 
proportional to $L^{-4}$,  $m_{max}=5$, the
parameterization~(\ref{favoritew4}), $|\mu| \le 1$,  and $L_{min}=24$
we get $\chi^2/$DOF $=1.052$ corresponding to $p=0.168$. The estimates
of the parameters are
$\lambda^*=5.14(4)$, $c=-1.17(17)$, and $d=0.07(1)$.

\item
For the Ansatz~(\ref{R3master}) without a correction of $U_4$ 
proportional to $L^{-2.023}$, no correction proportional to $L^{-4}$,
 $m_{max}=5$, the
parameterization~(\ref{favoritew4x}), $|\mu| \le 1$,  and $L_{min}=24$
we get $\chi^2/$DOF $=1.051$ corresponding to $p=0.171$. The estimates
of the parameters are
$\lambda^*=5.15(4)$, $c=-1.24(18)$, $d=0.05(3)$, and $e=0.07(6)$.

\item
For the Ansatz~(\ref{R3master}) with a correction of $U_4$ 
proportional to $L^{-2.023}$, no correction proportional to $L^{-4}$,
$m_{max}=5$, the 
parameterization~(\ref{favoritew4x}), $|\mu| \le 1$,  and $L_{min}=16$
we get $\chi^2/$DOF $=1.044$ corresponding to $p=0.159$. The estimates
of the parameters are
$\lambda^*=5.13(6)$, $c=-0.78(7)$, $d=0.05(1)$, and $e=0.04(3)$.

\item
For the Ansatz~(\ref{R3master}) without a correction of $U_4$
proportional to $L^{-2.023}$, but a correction 
proportional to $L^{-4}$ for all dimensionless quantities,  $m_{max}=5$, the
parameterization~(\ref{favoritew4x}), $|\mu| \le 1$,  and $L_{min}=12$
we get $\chi^2/$DOF $=1.041$ corresponding to $p=0.156$. The estimates
of the parameters are
$\lambda^*=5.096(20)$, $c=-0.78(4)$, $d=0.05(1)$, and $e=0.04(2)$.
\end{itemize}

In summary, also taking into account fits not explicitly given above, 
we find values of $\lambda^*$ that are consistent with 
$\lambda^*=5.17(11)$ obtained in Ref. \cite{myIco}. Furthermore 
$-1.5 \lessapprox c \lessapprox -0.7$, where the smaller values of $c$ 
are correlated with larger values of $\lambda^*$. There is only a small
asymmetry in $\mu$, corresponding to small values of $d$. 
These findings are consistent with the results for $\lambda^*(\mu)$, 
which are summarized in table \ref{N3ls}.

Finally in tables \ref{N3betac1} and \ref{N3betac2}
 we give the results for $\beta_c$ which
are based on the four fits which are explicitly discussed above.
Given the large number of pairs $(\lambda,\mu)$ we simulated
at, we used an automated procedure to obtain the central value and
its error, similar to the analysis for $N=4$ above.
These results might be used to bias the analysis of high temperature (HT) 
series expansions or in future Monte Carlo studies of the model.

\begin{table}
\caption{\sl \label{N3betac1}
Numerical results for the inverse critical temperature $\beta_c$
for the pairs of $(\lambda,\mu)$ we simulated at for $N=3$. For a
discussion see the text.
}
\begin{center}
\begin{tabular}{crl}
\hline
 \mc{1}{c}{$\lambda$}  &  \mc{1}{c}{$\mu$} &  \mc{1}{c}{$\beta_c$} \\
\hline
   4.3 &   -1.00 &   0.6773490(14) \\
   4.5 &   -1.00 &   0.67841424(96) \\
   4.5 &   -0.50 &   0.68439865(37) \\
   4.5 &   -0.30 &   0.68559135(40) \\
   4.5 &   -0.15 &   0.68607863(25) \\
   4.5 &    0.15 &   0.68608422(22) \\
   4.5 &    0.25 &   0.68581697(32) \\
   4.5 &    0.30 &   0.68563593(23) \\
   4.5 &    0.35 &   0.68542342(26) \\
   4.5 &    0.50 &   0.68460540(23) \\
   4.5 &    1.00 &   0.68006803(59) \\
\hline
\end{tabular}
\end{center}
\end{table}

\begin{table}
\caption{\sl \label{N3betac2}
Continuation of table \ref{N3betac1}.
}
\begin{center}
\begin{tabular}{crl}
\hline
 \mc{1}{c}{$\lambda$}  &  \mc{1}{c}{$\mu$} &  \mc{1}{c}{$\beta_c$} \\
\hline
   4.7 &   -0.70 &   0.68329292(59) \\
   4.7 &    0.70 &   0.68382797(22) \\
   4.8 &   -0.50 &   0.68536469(28) \\
   4.8 &   -0.30 &   0.68647819(18) \\
   4.8 &    0.20 &   0.68682876(15) \\
   4.8 &    0.30 &   0.68651908(10) \\
   4.8 &    0.40 &   0.68609278(13) \\
   4.8 &    0.50 &   0.68555428(13) \\ 
   5.0 &   -0.30 &   0.68698276(13) \\
   5.0 &   -0.10 &   0.68749850(13) \\ 
   5.0 &    0.00 &   0.68756126(8) \\
   5.0 &    0.10 &   0.68749982(12) \\
   5.0 &    0.20 &   0.68731855(11) \\ 
   5.0 &    0.25 &   0.68718435(9) \\
   5.0 &    0.30 &   0.68702161(6) \\
   5.0 &    0.40 &   0.68661260(11) \\
   5.2 &   -0.50 &   0.68640695(35) \\
   5.2 &   -0.30 &   0.68742991(35) \\
   5.2 &   -0.10 &   0.68792511(14) \\ 
   5.2 &    0.00 &   0.68798524(8) \\
   5.2 &    0.05 &   0.68797037(15) \\
   5.2 &    0.10 &   0.68792634(12) \\
   5.2 &    0.20 &   0.68775221(10) \\
   5.2 &    0.30 &   0.68746677(11) \\
   5.2 &    0.40 &   0.68707374(14) \\
   5.2 &    0.50 &   0.68657694(16) \\
   5.2 &    0.70 &   0.68528562(32) \\ 
\hline
\end{tabular}
\end{center}
\end{table}

\subsubsection{$U_C$ at $Z_a/Z_p=0.19477$}
\begin{figure}
\begin{center}
\includegraphics[width=14.5cm]{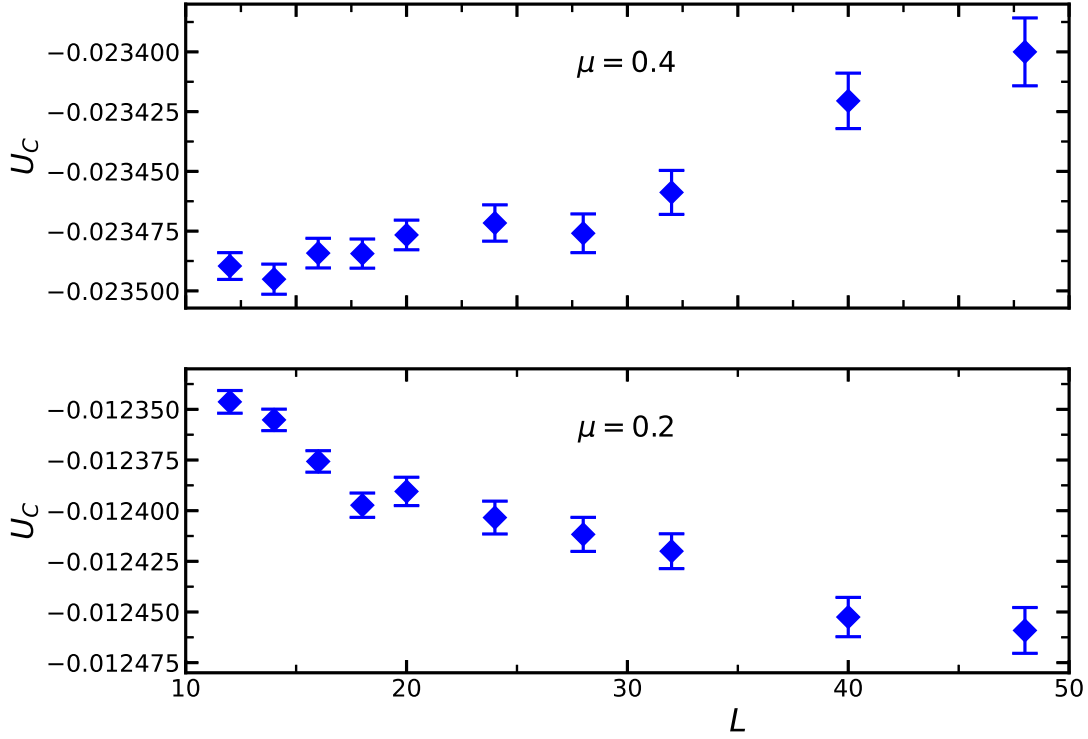}
\caption{\label{barUCplot}
We plot $U_C$ at $Z_a/Z_p=0.19477$ versus the linear lattice size $L$
for $\lambda=5.0$ and $N=3$. In the upper part we give numerical results for 
$\mu=0.4$ and in the lower part the estimates for $\mu=0.2$.
}
\end{center}
\end{figure}

In a complementary analysis we considered $U_C$ at $Z_a/Z_p=0.19477$. 
To get a first impression of its behavior, we plot
$\bar{U}_C$ as a function of the linear lattice size $L$ for 
$(\lambda,\mu)=(5.0,0.2)$ and $(5.0,0.4)$ in Fig. \ref{barUCplot}.  
We find that  $\bar{U}_C$ is slowly decreasing with increasing 
lattice size for $\mu=0.2$, while it is increasing for  $\mu=0.4$.
We notice that high statistical accuracy is needed to detect this behavior.
We expect that $0.2 < \mu^* < 0.4$. For $\mu<0$, $\bar{U}_C$ is positive and it 
is increasing with increasing lattice size, throughout.

Let us analyze $\bar{U}_C$ quantitatively.
We performed joint fits for different values of $\mu$ and a single value
of $\lambda$, either $\lambda=4.8$, $5.0$ or $\lambda=5.2$. 

First we performed fits by using the Ansatz~(\ref{UCbasic})
\begin{equation}
\label{UCbasicA}
\bar{U}_C(\mu,\lambda,L) =
 \frac{\bar{U}_C^*}{1 + q \left(\frac{\mu^*}{\mu} -1 \right) L^{-y} }
  \;,
\end{equation}
where we simply have replaced $U_C$ by $\bar{U}_C$.
Next we introduce a quadratic correction in $\mu$:
\begin{equation}
\label{UCbasicQ2}
\bar{U}_C(\mu,\lambda,L) =
 \frac{\bar{U}_C^*}{1 + q \left(\frac{\bar{\mu}}{\mu+s \mu^2} -1 \right) L^{-y} } 
  \;,
\end{equation}
where now
\begin{equation}
\mu^* = \frac{\sqrt{1+4 \bar{\mu} s }-1}{2 s}
\end{equation}
or a correction proportional to $L^{-2+\eta}$ 
\begin{equation}
\label{UCbasicX}
\bar{U}_C(\mu,\lambda,L) =
 \frac{\bar{U}_C^*}{1 + q \left(\frac{\mu^*}{\mu} -1 \right) L^{-y} } 
\left(1 + c L^{-2+\eta}  \right) \;.
\end{equation}
and both types of corrections 
\begin{equation}
\label{UCbasicX2}
\bar{U}_C(\mu,\lambda,L) =
\frac{\bar{U}_C^*}{1 + q \left(\frac{\bar{\mu}}{\mu+s \mu^2} -1 \right) 
L^{-y} }
\left(1 + c L^{-2+\eta}  \right)
 \;.
\end{equation}

We find that acceptable fits can only be obtained by restricting the 
range of the parameter: $|\mu| \le 0.4$.  
In the case of $\lambda=5.0$ we get, by
using the Ansatz~(\ref{UCbasicX2}) a
$\chi^2/$DOF $=1.01$ for $L_{min}=12$ and $\chi^2/$DOF slightly smaller than 
one for larger $L_{min}$. 
We obtain $\bar{U}_C^* = -0.0176(3)$, $-0.0172(4)$, $-0.0171(5)$, $-0.0168(6)$, 
and $-0.0166(7)$ for $L_{min}=12$, $14$, $16$, $18$, and $20$, respectively. 
Furthermore $y=0.0149(3)$, $0.0146(4)$, $0.0144(5)$, $0.0142(5)$, and 
$0.0141(6)$ for the same values of $L_{min}$ as above. Note that
$y=Y_4=\omega_2$ in the approximation used here.
For the fixed point value of the parameter $\mu$ we get
$\mu^*=0.290(6)$, $0.283(8)$, $0.283(9)$, $0.277(11)$, and $0.273(12)$.

Comparing with the results obtained by using the other Ans\"atze and 
$\lambda=4.8$ and $5.2$ we arrive at the final results 
\begin{equation}
 Y_4 = 0.0144(15) 
\end{equation}
and
\begin{equation}
\label{UCstar}
 \bar{U}_C^* =  -0.017(2)  \;.
\end{equation}
The error bars are chosen such that the estimates of different acceptable fits
are covered.
For $\lambda=5.0$ we conclude $\mu^*=0.28(2)$. The estimates for 
$\lambda=4.8$ and $5.2$ are the same within errors. 

We have repeated the analysis for $U_C$ at 
$Z_a/Z_p-0.64 U_C^2+ 2.1 U_C^3 =0.19477$. We arrive at very similar results.
(Revised version: See remark in Sec. \ref{UCflow}).

Putting things together, the improved model for the cubic fixed point is given
by $(\lambda,\mu)^*_{cubic} = (4.99(11),0.28(2))$, where we obtain the 
value of $\lambda$ by interpolating the estimates for $\mu=0.2$ and $0.3$
given in table \ref{N3ls}.

We performed fits with Ans\"atze that combine Eq.~(\ref{R3master}) with 
the Ans\"atze for $U_C$ discussed in this subsection.  The results are 
fully consistent with those given above. Therefore we abstain
from a discussion.

\subsubsection{Generic Ansatz for the dimensionless quantities in the 
neighborhood of the cubic fixed point}
Finally we performed fits, similar to the case $N=4$, with a generic Ansatz,
not exploiting the vicinity of the $O(3)$ symmetric fixed point.
In order to get an
acceptable $\chi^2/$DOF, using the parameterization~(\ref{para_u}), the values
of $(\lambda,\mu)$ have to be restricted to a close neighborhood of
$(\lambda,\mu)^*$. Here we only included data with $0.2 \le \mu \le 0.4 $
for $\lambda=4.8$, $5.0$, and $5.2$.
We are aiming at estimates of the fixed point values of the 
dimensional quantities $R_i^*$ and  $(\lambda,\mu)^*$. 

First we take $\omega_2$ as free parameter in the Ansatz~(\ref{RGansatz1}), 
while we fix $\omega_1=0.759$
\cite{myIco}. We get an acceptable goodness of the fit starting from
$L_{min}=18$. We get $\omega_2=0.0150(13)$, $0.0137(17)$, $0.0179(23)$, 
and $0.021(3)$ for $L_{min}=12$, $14$, $16$, and $18$.
Going to larger values of $L_{min}$, the statistical error is rapidly 
increasing. Therefore we performed
fits fixing $\omega_2=0.0143$. In this case, we also get an acceptable 
goodness of the fit starting from $L_{min}=18$. As a check, we also 
performed fits using $\omega_{2}=0.013$, taking into account possible 
deviations of $\omega_{2}$ from $Y_4$ of the $O(3)$-invariant fixed point. 

We arrive at $(\lambda,\mu)^*_{cubic} = (4.98(10),0.30(3))$ covering
results for $L_{min}=18$ up to $24$.  The differences of results for 
$\omega_2=0.013$ and $0.0143$ are clearly smaller than the error bars quoted.

Furthermore we get
\begin{eqnarray}
(Z_a/Z_p)^*_{cubic}     &=& 0.19453(5) \\
(\xi_{2nd}/L)^*_{cubic} &=& 0.56451(7) \\
U_{4,cubic}^* &=&  1.13972(6) \\
U_{C,cubic}^* &=&  -0.0181(12) 
\end{eqnarray}
for the cubic fixed point. Also here, the estimates for $L_{min}=18$ up 
to $L_{min}=24$ are covered and the difference of results for
$\omega_2=0.013$ and $0.0143$ are clearly smaller than the error bars quoted.

The estimates of $(Z_a/Z_p)^*_{cubic}$, $(\xi_{2nd}/L)^*_{cubic}$ and
$U^*_{4,cubic}$ differ only
slightly from the values for the $O(3)$ symmetric fixed point. However
the differences are clearly larger than the error estimates.

\subsubsection{Flow equation for $\bar{U}_C$}
\label{UCflow}
Finally we consider the dimensionless quantity $U_C$ itself as coupling.
In order to stay at criticality we take it at
$Z_a/Z_p-0.64 U_C^2+ 2.1 U_C^3 =0.19477$. 

(Remark revised version: 
$Z_a/Z_p-0.64 U_C^2+ 2.1 U_C^3$ should be 
$Z_a/Z_p+0.64 U_C^2- 2.1 U_C^3$. However even with the wrong sign it is a 
dimensionless quantity that can be fixed to a certain value. Errors
in the final result might become larger due to larger corrections. However,
we have checked that  replacing $-0.64 U_C^2+ 2.1 U_C^3$ by
$+0.64 U_C^2- 2.1 U_C^3$ changes virtually nothing in the final results.
Therefore we did not update the discussion of the analysis of the data
below.)

Furthermore
we stay, at the level of our numerical precision, on the line $\lambda^*(\mu)$. 

We determine 
\begin{equation}
u = \frac{1}{\bar{U}_C}  \frac{\mbox{d} \bar{U}_C}{\mbox{d}l}  \;,
\end{equation}
where $l=\ln L$, by fitting the data for fixed $(\lambda,\mu)$ by using 
the Ansatz
\begin{equation}
\label{mostsimple}
\bar{U}_C(\lambda,\mu,L)  = a L^{u}
\end{equation}
for some range $L_{min} \le L \le  L_{max}$. As argument of $u$ we take
$[\bar{U}_C(L_{min}) + \bar{U}_C(L_{max})]/2$. 
The approximation~(\ref{mostsimple}) 
relies on the fact that $\bar{U}_C$ varies only little
in the range of linear lattice sizes considered.
In order to check the effect of
subleading corrections, we consider different ranges 
$L_{min} \le L \le L_{max}$.  For $L_{min}=32$ the maximal lattice size
$L_{max}$ is determined by the largest lattice 
size that we have simulated.  For $L_{min}=16$ and $24$, we reduce 
$L_{max}$ by the corresponding factor with respect to $L_{min}=32$.
Finally we used the Ansatz
\begin{equation}
\label{mostsimple2}
\bar{U}_C(\lambda,\mu,L)  = a L^{u} \; (1+ c L^{-2}) 
\end{equation}
with $L_{min}=12$ and $L_{max}$ given by the largest lattice size
simulated. In our analysis, we took into account the data for 
$(\lambda,\mu)=(4.3,-1)$, $(4.5,-1)$, $(4.7,-0.7)$, $(4.8,-0.5)$, 
$(5.0,-0.3)$, 
$(5.0,-0.1)$, $(5.2,-0.1)$, $(5.0, 0.1)$, $(5.2, 0.1)$,
$(5.0,0.2)$, $(5.2,0.2)$, $(5.0,0.25)$, $(5.0,0.3)$, 
$(5.0,0.4)$, $(4.8,0.5)$, $(4.7,0.7)$, and $(4.5,1.0)$.

We fit the estimates of $u$ by using the Ansatz
\begin{equation}
\label{uAn1}
 u(\bar{U}_C) = a + b \bar{U}_C + c \bar{U}_C^2 + d \bar{U}_C^3
\end{equation}
and as check
\begin{equation}
\label{uAn2}
 u(\bar{U}_C) = a + b \bar{U}_C + c \bar{U}_C^2 \;.
\end{equation}
It turns out that the Ansatz~(\ref{uAn2}) gives quite large $\chi^2/$DOF, 
when all data are fitted, while Ansatz~(\ref{uAn1}) results in 
an acceptable $\chi^2/$DOF.  Excluding the data for $|\mu|=1$, also 
Ansatz~(\ref{uAn2}) gives acceptable values of $\chi^2/$DOF.

\begin{figure}
\begin{center}
\includegraphics[width=14.5cm]{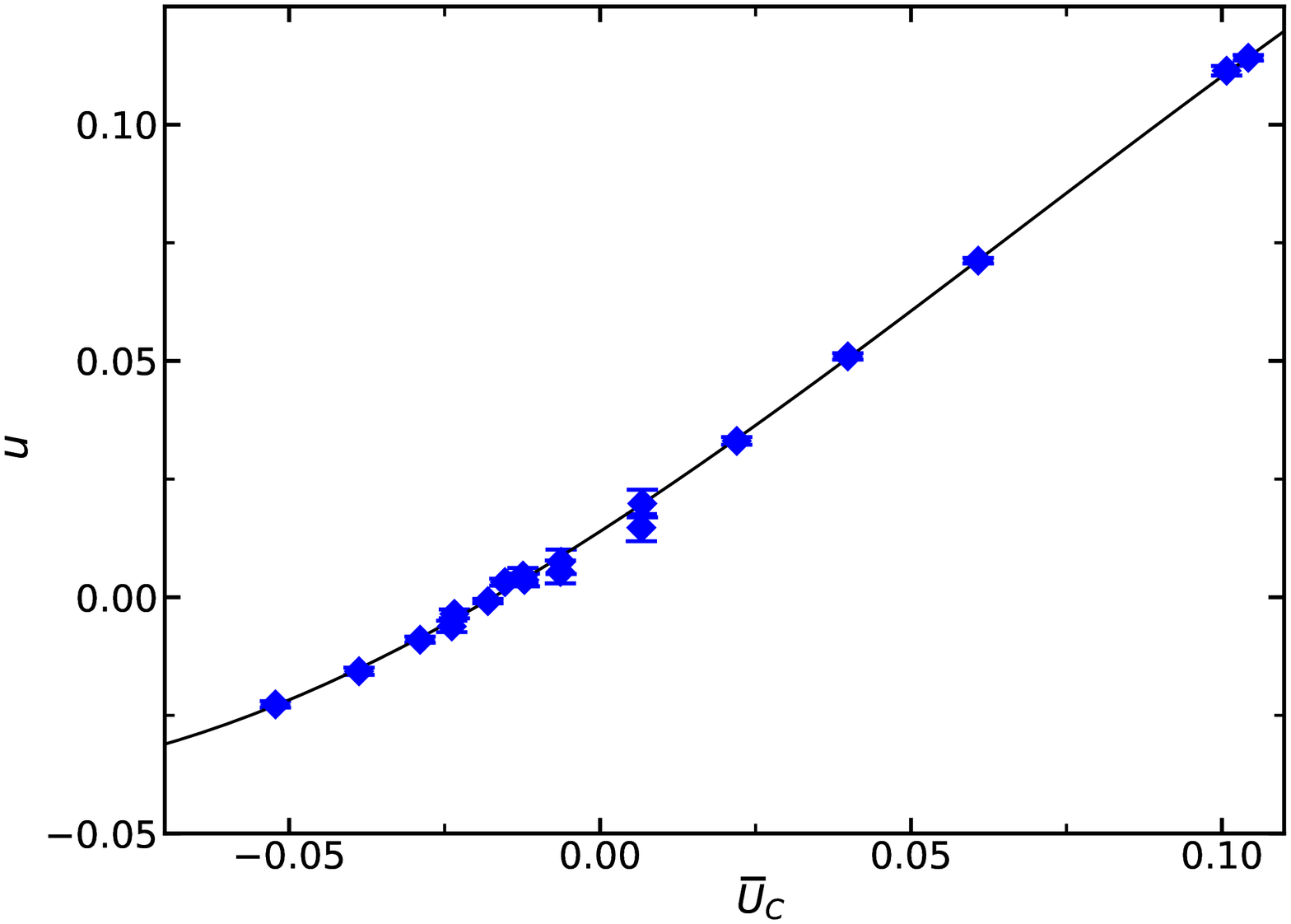}
\caption{\label{uplot}
We plot $u \approx \frac{1}{\bar{U}_C} \frac{\mbox{d} \bar{U}_C}{\mbox{d}l}$ 
for $N=3$ as a function of $\bar{U}_c$. Here we give the data obtained by
using the Ansatz~(\ref{mostsimple2}) and $L_{min}=12$.
The line gives the result of the fit with the Ansatz~(\ref{uAn2}).
}
\end{center}
\end{figure}

In Fig.~(\ref{uplot}) we plot the numerical estimates of $u(\bar{U}_c)$ 
obtained 
by using the Ansatz~(\ref{mostsimple2}) with $L_{min}=12$. 
The line corresponds to the fit of the data by using the 
Ansatz~(\ref{uAn1}).
The relative error of the data for $|\mu| \le 0.2$ is large. These data
contribute little to the final result.

In table \ref{Fitu} we  summarize the numerical results. In addition
to the estimates of the parameters of the Ans\"atze~(\ref{uAn1},\ref{uAn2})
we give the zero $\bar{U}_C^*$ of $u$ and the correction exponent $\omega_2$ 
at this zero. These are computed numerically for the given estimates
of $a$, $b$, $c$ and $d$.

\begin{table}
\caption{\sl \label{Fitu}
Results of fitting $u(\bar{U}_c)$ by using the Ansatz~(\ref{uAn1}) or 
~(\ref{uAn2}). The estimates of $u$ are obtained by using the 
Ansatz~(\ref{mostsimple}) or (\ref{mostsimple2}) using the minimal lattice
size $L_{min}$. The corresponding maximal lattice size is given in the text.
$a$, $b$, $c$ and $d$ are the 
parameters of the Ans\"atze~(\ref{uAn1},\ref{uAn2}). 
If no value for $d$ is given, Ansatz~(\ref{uAn2}) is used.
Otherwise the data are fitted by using the Ansatz~(\ref{uAn1}).
$\bar{U}_C^*$ is the zero 
of $u$ that is computed numerically and $\omega_2$ the correction exponent 
at this zero.  (Remark revised version: by mistake we have computed the 
statistical error of $a+\omega_2$ instead of $a-\omega_2$. The error
of $a-\omega_2$ is smaller than that of $a+\omega_2$. In the table below, 
still the wrong error is quoted.)
}
\begin{center}
\begin{tabular}{ccccccccc}
\hline
Ansatz & $L_{min}$  & a  &  b  & c & d & $\bar{U}_C^*$ & $\omega_2$ & $a-\omega_2$  \\
\hline
\ref{mostsimple} & 16 & 0.01558(22) & 0.836(6) & 2.00(13) & -11.9(1.4) & -0.0197(3) &  0.01463(25) & 0.00096(32)  \\
\ref{mostsimple} & 24 & 0.01398(41)& 0.843(10)& 2.64(30) & -17.7(2.9) & -0.0177(5) &  0.01296(46) & 0.00102(47) \\
\ref{mostsimple} & 32 & 0.01521(48)  & 0.826(12)&  0.94(16)&    -       &-0.0188(5) &  0.01488(51) & 0.00033(53) \\
\ref{mostsimple} & 32 & 0.01429(55) & 0.851(14)&  2.06(35)& -12.2(3.4) & -0.0176(6)
&  0.01352(60)& 0.00077(78) \\
\ref{mostsimple2} & 12 & 0.01465(30) & 0.848(9) & 1.20(16) &     -      & -0.0177(3) &  0.01427(33) & 0.00038(34) \\
\ref{mostsimple2} & 12 & 0.01392(31) & 0.850(8) & 2.20(20) & -10.7(2.1) & -0.0172(4) &  0.01316(34) & 0.00076(43)  \\
\hline
\end{tabular}
\end{center}
\end{table}

The results obtained by using the Ansatz~(\ref{mostsimple}) 
with $L_{min}=24$ and $32$ and those
obtained by using the Ansatz~(\ref{mostsimple2}) and $L_{min}=12$ are 
essentially consistent. Fitting $u(\bar{U}_c)$ by using the
 Ansatz~(\ref{uAn1}), the 
results for $a$ are slightly smaller than by using the Ansatz~(\ref{uAn2}).
Furthermore, the difference $a-\omega_2$ is smaller when fitting by using
the Ansatz~(\ref{uAn2}) than for Ansatz~(\ref{uAn1}). Giving preference to
the Ansatz~(\ref{uAn1}) and fitting all data, we arrive at the results
$Y_4=a=0.0142(6)$, $\omega_2=0.0133(8)$, and 
$\bar{U}_C^*=-0.0175(7)$.  Throughout the fits reported in table \ref{Fitu},
$\omega_2<Y_4$, and 
$Y_4-\omega_2<0.0015$ in the extreme case, taking into account the 
statistical error.
The estimates of $Y_4$ and $\bar{U}_C^*$ are consistent with those obtained in 
previous sections.

\subsection{The critical exponent $\eta$} 
(Remark revised version: Also here the wrong sign of corrections in $U_C$
is taken. However a reanalysis, taking the correct sign, or abstain
from using it, shows that the final result for $\eta$ is virtually uneffected.)

\begin{figure}
\begin{center}
\includegraphics[width=14.5cm]{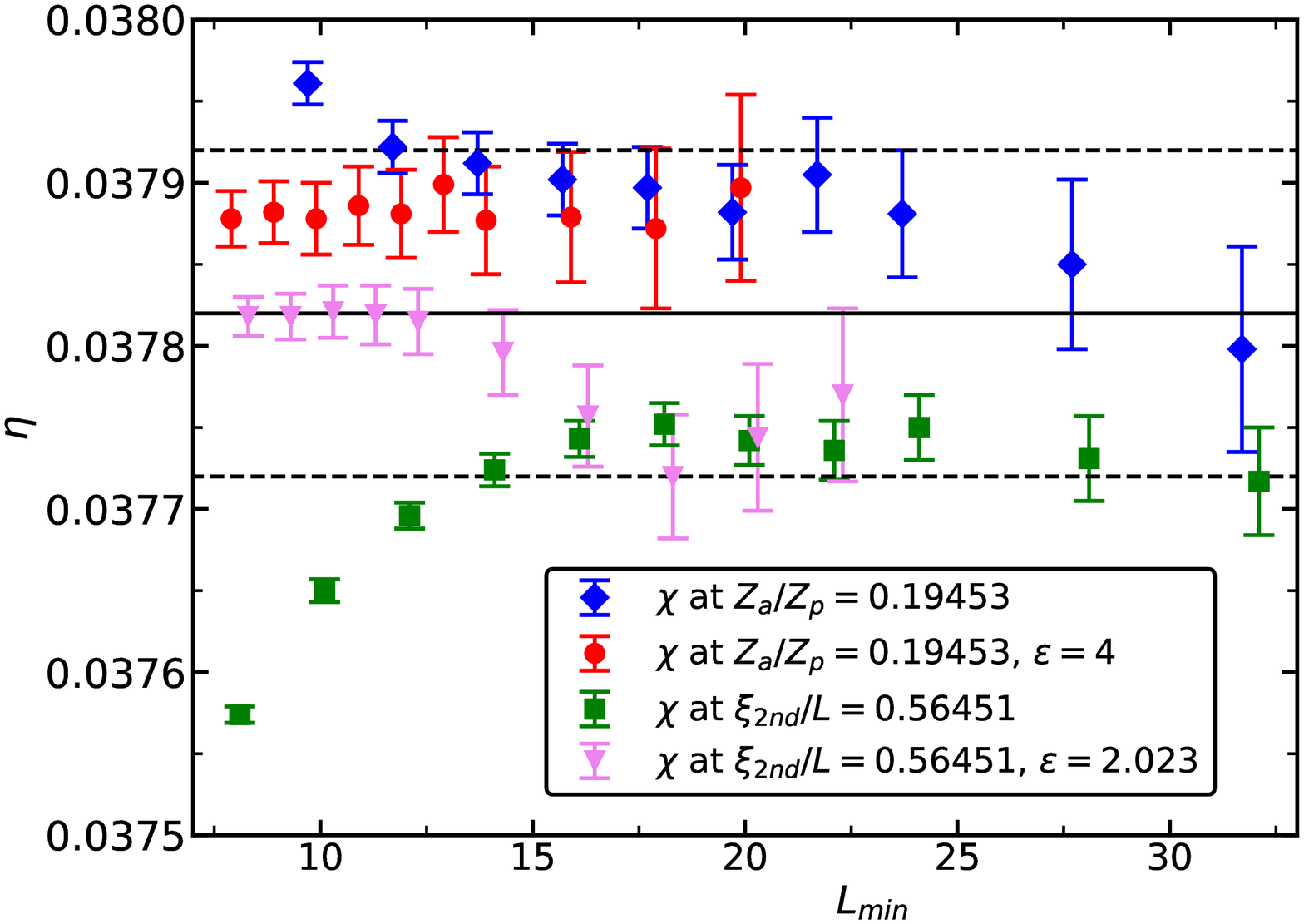}
\caption{\label{etaN3}
We give estimates of $\eta$ obtained by fitting the data for $\chi$ at 
$(\lambda,\mu)=(5.0,0.3)$ and $N=3$
by using the Ans\"atze~(\ref{chifit1},\ref{chifit2}). In the legend, 
for the Ansatz~(\ref{chifit2}), we give the value of the correction exponent
$\epsilon$.
Note that the values on the $x$-axis are slightly shifted to reduce overlap
of the symbols.
The solid line gives our preliminary estimate of
$\omega$, while the dashed lines indicate the error.
}
\end{center}
\end{figure}

Here we focus on the analysis of our data for $(\lambda,\mu)=(5,0.3)$. 
We analyze the magnetic susceptibility $\chi$ at
$Z_a/Z_p=0.19453$ or $\xi_{2nd}=0.56451$. 
We used the Ans\"atze~(\ref{chifit1},\ref{chifit2}) already used for $N=4$. 
Our estimates of $\eta$ are plotted in Fig. \ref{etaN3}. 
As our preliminary estimate we take $\eta=0.03782(10)$ that covers, for some 
range of $L_{min}$, the results obtained from all four fits.

In order to estimate the dependence of the result on $\lambda$, we analyze
the data for $\lambda=4.8$ and $5.2$. Assuming that subleading corrections 
to scaling are very similar for these values of $\lambda$ we compare
fits with small $L_{min}$, where the statistical error is small.  We find, 
consistently for both Ans\"atze~(\ref{chifit1},\ref{chifit2}) and fixing 
$Z_a/Z_p=0.19453$ or $\xi_{2nd} =0.56451$ that the estimates of $\eta$ for 
 $\lambda=4.8$  are larger by about $0.0001$ than for $\lambda=5.2$.
In the analysis of the data for $\lambda=4.8$ and $5.2$ smaller lattices 
are included than for $\lambda=5.0$. Therefore the effect
of corrections proportional to $L^{-\omega_1}$ should be smaller. 
Given the accuracy of $\lambda^*$ for the cubic
fixed point we arrive at our final estimate 
\begin{equation}
\label{finaleta3}
 \eta=0.03782(13) \;.
\end{equation}
This estimate is within the errors consistent with that obtained 
in Ref. \cite{myIco} for the $O(3)$-invariant fixed point: 
$\eta_{O(3)}=0.03784(5)$.  Therefore, assuming that the estimate of $\eta$ 
is monotonic in the scaling field of the cubic perturbation in the range that
we consider here, we do not add an additional error due to the uncertainty
of $\mu^*$.

\subsection{The critical exponent $\nu$}
(Remark revised version: Also here the wrong sign of corrections in $U_C$
is taken. A reanalysis shows that the final result for the critical exponent
$\nu$ at the cubic fixed point is uneffected.)

\begin{figure}
\begin{center}
\includegraphics[width=14.5cm]{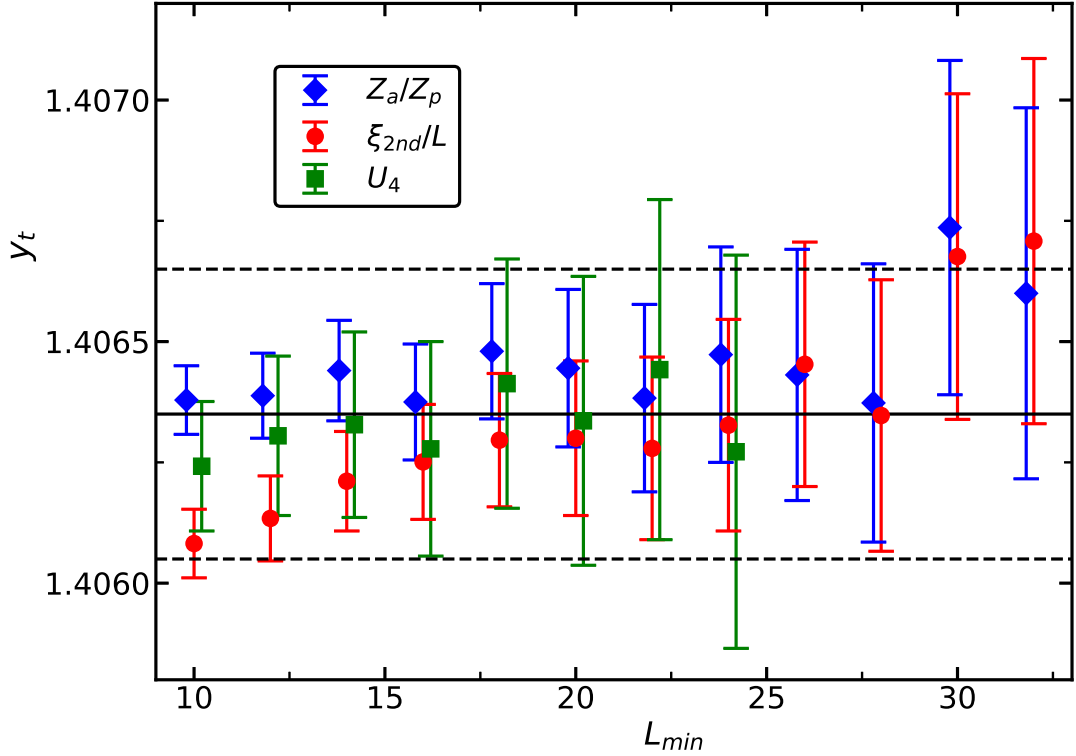}
\caption{\label{ytN3}
We give estimates of $y_t$ obtained by fitting the data for the slopes of
$Z_a/Z_p-0.64 U_C^2+ 2.1 U_C^3$, ..., at $(\lambda,\mu)=(5.0,0.3)$ and $N=3$
by using the Ansatz~(\ref{slopeAnsatz}).  
Note that the values on the $x$-axis are slightly shifted to reduce overlap
of the symbols. The solid line gives our
preliminary estimate of $y_{t}$ and the dashed lines indicate the error.
}
\end{center}
\end{figure}

We  have analyzed the slopes of dimensionless quantities
$Z_a/Z_p-0.64 U_C^2+ 2.1 U_C^3$, $\xi_{2nd}/L + 1.34 U_C^2 -3.4 U_C^3$, 
and $U_4-1.25 U_C^2- 3.0 U_C^3$  at $Z_a/Z_p-0.64 U_C^2+ 2.1 U_C^3=0.19477$
that stay approximately constant on the line $\lambda^*(\mu)$ at criticality.
Below we denote these quantities by $Z_a/Z_p+ ...$, $\xi_{2nd}/L + ...$, 
and $U_4 + ...$ for simplicity.
We performed fits with the Ansatz~(\ref{slopeAnsatz}).
The resulting estimates of $y_t$  are plotted in Fig.~\ref{ytN3}. 
As our preliminary 
estimate we take $y_t =1.40635(30)$.  In order to estimate the effect
of corrections proportional to $L^{-\omega_1}$, we analyze 
ratios
\begin{equation}
 r_{S,i}(L) = \frac{S_{\lambda=5.2,i}(L)}{S_{\lambda=4.8,i}(L)} \; ,
\end{equation}
where $i$ indicates which dimensionless quantity is taken. We expect 
that subleading corrections approximately cancel. Therefore we analyze
these ratios with the simple Ansatz
\begin{equation}
r_{S,i}(L) =a L^{\Delta y_{t}} \; .
\end{equation}
The estimate for $L_{min}=16$ is $\Delta y_{t}=-0.00044(10)$, 
$-0.00030(10)$ and $0.00021(19)$ for the slopes of $Z_a/Z_p+ ...$, 
$\xi_{2nd}/L + ...$ and
$U_4+ ...$, respectively. 
Since the difference in $\lambda$ is about 4 times as large 
as the uncertainty of $\lambda$ in $(\lambda,\mu)^*$,
we conclude that the error of $y_{t}$ due to the 
uncertainty of $\lambda$ in $(\lambda,\mu)^*$ is about $0.0001$.  Finally
we analyzed the ratios
\begin{equation}
 r_{S,i}(L) = \frac{S_{\mu=0.3,i}(L)}{S_{\mu=0.25,i}(L)} \; 
\end{equation}
for $\lambda=5.0$. Here we get 
$\Delta y_{t}=0.00029(9)$,  $0.00019(8)$, and $0.00022(16)$ for 
or $Z_a/Z_p+ ...$, $\xi_{2nd}/L+ ...$ and $U_4+ ...$, respectively. Taking 
the estimate $\mu^* = 0.28(2)$  
we arrive at the final estimate
\begin{equation}
 y_{t,cubic} = 1.40625(50) \;,
\end{equation}
which can be compared with $y_{t,O(3)}=1.4052(2)$ \cite{myIco}.

\subsection{Difference between critical exponents for the $O(3)$-invariant and
the cubic fixed point}
\label{ratio_slope}
(Remark revised version: Also here the wrong sign of corrections in $U_C$
is taken. A reanalysis shows that the final results for the critical exponents
$\nu$ and $\eta$ at the cubic fixed point are uneffected. With the correct
sign, the estimates of $c_2$ given in table \ref{Fityt} are consistant 
for the three different dimensionless quantities, with $c_2 \approx 3.8$, 
while $c_3$ still assumes different values for different dimensionless 
quantities.)

Based on the expectation that corrections to scaling are similar for the 
improved models for the $O(3)$-invariant and the cubic fixed point, we
study ratios of magnetic susceptibilities and slopes at criticality. 
In addition to $(\lambda,\mu)^*$  we analyze data for pairs $(\lambda,\mu)$
that are approximately on the line $\lambda^*(\mu)$. 
To stay critical we take the quantities at either
$Z_a/Z_p-0.64 U_C^2+ 2.1 U_C^3 =0.19477$ or 
$\xi_{2nd}/L + 1.34 U_C^2 -3.4 U_C^3 = 0.56404$.
We evaluate ratios of slopes $S$
\begin{equation}
 r_{S,i}(L) = \frac{S_{Cubic,i}(L)}{S_{O(3),i}(L)} \; ,
\end{equation}
where $i$ indicates which dimensionless quantity is taken.
We analyze these ratios by fitting with the simple Ansatz
\begin{equation}
\label{ratiofit1}
r_{S,i}(L) =a  L^{y_{t,Cubic}-y_{t,O(3)}}
\end{equation}
or as check
\begin{equation}
\label{ratiofit2}
r_{S,i}(L) =a L^{y_{t,Cubic}-y_{t,O(3)}}  (1+c L^{-2}) \;.
\end{equation}
In the case of the magnetic susceptibilities we use analogous Ans\"atze.

Let us first analyze the magnetic susceptibility.
We only discuss $\chi$ at $\xi_{2nd}/L  + 1.34 U_C^2 -3.4 U_C^3 = 0.56404$,
since the statistical
error of $\chi$ at $\xi_{2nd}/L  + 1.34 U_C^2 -3.4 U_C^3 = 0.56404$ is clearly
smaller than at $Z_a/Z_p-0.64 U_C^2+ 2.1 U_C^3 =0.19477$. 

For the ratio of the susceptibility at $(\lambda,\mu)=(5.0,0.3)$ and
$(5.2,0)$ we get $\Delta \eta = 0.00004(3)$ taking into
account both the analogues of the Ans\"atze~(\ref{ratiofit1},\ref{ratiofit2}).
As check, we computed the ratio for $(\lambda,\mu)=(5.0,0.3)$ and
$(5.0,0)$. We get $\Delta \eta = 0.00002(3)$.

A $\Delta \eta$ that is clearly different 
from zero we only get for larger values of $|\mu|$. For example for
$(\lambda,\mu)=(4.7,0.7)$ and $(5.2,0)$ we get 
$\Delta \eta = 0.00024(5)$ and for 
$(\lambda,\mu)=(4.8,-0.5)$ and $(\lambda,\mu)=(5.2,0)$ we get
$\Delta \eta = - 0.00040(5)$.  We regard the estimates obtained from
the susceptibility at $(\lambda,\mu)=(5.0,0.3)$ and $(5.2,0)$
or $(5.0,0)$
as bound for the difference between the cubic and the $O(3)$-invariant
fixed point. Therefore
\begin{equation}
\label{etadiff}
-0.00001  \lessapprox \eta_{cubic} - \eta_{O(3)} \lessapprox 0.00007 \;,
\end{equation}
which is more strict than the difference of our result~(\ref{finaleta3}) 
and the estimate of $\eta_{O(3)}$ of Ref. \cite{myIco}. 

\begin{figure}
\begin{center}
\includegraphics[width=14.5cm]{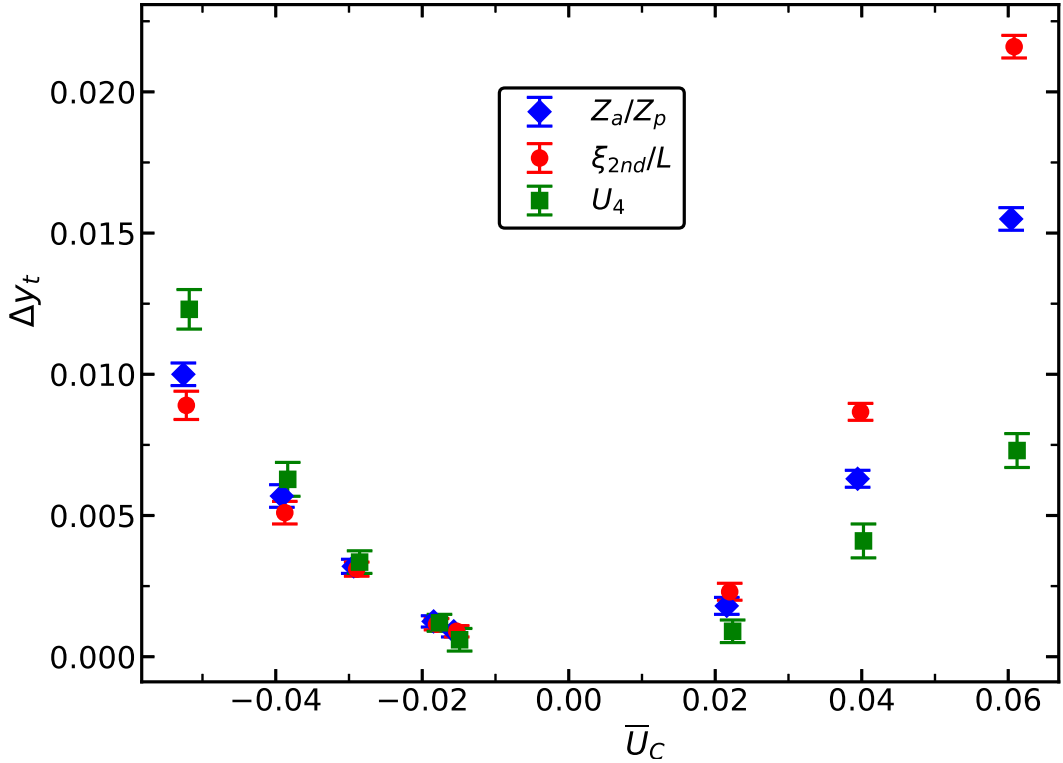}
\caption{\label{Dyt3}
We plot $\Delta y_t$ obtained by fitting ratios of slopes for 
$N=3$ by using the Ans\"atze~(\ref{ratiofit1},\ref{ratiofit2}) as a function
of $\bar{U}_C$.
For a discussion see the text.
}
\end{center}
\end{figure}

Finally we  study ratios of slopes for $(\lambda,\mu)=(5.2,0)$ 
and several pairs $(\lambda,\mu)$ that approximate $\lambda^*(\mu)$. 
Our estimates are given in Fig.~(\ref{Dyt3}) as a function of $\bar{U}_C$, 
where $\bar{U}_C=[\bar{U}_C(L_{max})+\bar{U}_C(L_{min})]/2$, 
similar to section \ref{UCflow}.

We have fitted the estimates with the Ansatz
\begin{equation}
\label{DeltaAnsatz}
 \Delta y_t = c_2 \bar{U}_C^2 + c_3 \bar{U}_C^3 \;.
\end{equation} 
The results for the coefficients are given in table \ref{Fityt}.
Plugging in the estimate $\bar{U}_C^*=-0.0175(7)$, sec. \ref{UCflow},
we arrive at $\Delta y_t = 0.00117(13)$, $0.00123(13)$, and $0.00114(15)$ for
$Z_a/Z_p + ...$, $\xi_{2nd}/L + ...$, and $U_4 + ...$, respectively.
As our final estimate we quote
\begin{equation}
 y_{t,Cubic}-y_{t,O(3)} = 0.0012(2) \;,
\end{equation}
where the error is dominated by the uncertainty of $U_C^*$. 
This result translates to
\begin{equation}
\label{nudiff}
 \nu_{Cubic}-\nu_{O(3)} =-0.00061(10) \;.
\end{equation}

\begin{table}
\caption{\sl \label{Fityt}
Coefficients of $\Delta y_t = c_2 \bar{U}_C^2 + c_3 \bar{U}_C^3$ for the slopes of
three different dimensionless quantities for $N=3$. 
For a discussion see the text.
}
\begin{center}
\begin{tabular}{clr}
\hline
  $R$         &  \mc{1}{c}{$c_2$}   &  \mc{1}{c}{$c_3$}  \\
\hline
$Z_a/Z_p + ...$     & 3.90(8)  &  4.3(1.5)  \\
$\xi_{2nd}/L + ...$ & 4.42(8)  &  23.6(1.6) \\
$U_4 + ...$         & 3.32(13) & -22.2(2.4) \\
\hline
\end{tabular}
\end{center}
\end{table}

\section{Comparison with results given in the literature
}
In the literature,
information on the cubic fixed point stems mainly from field theoretic
methods. The $\epsilon$-expansion has been computed up to 5-loop in 
Ref. \cite{Frohlinde} and has recently been extended to 6-loop 
\cite{epsilon6}. The perturbative expansion in $d=3$ fixed has been 
computed up to 6-loop in Ref. \cite{Carmona}.
The numerical values
obtained for the critical exponents vary with the resummation scheme that is
used. For example, the 6-loop $\epsilon$-expansion for $N=3$ has been 
resummed in Ref. \cite{epsilon6} by using the Pad\'e approximation 
and alternatively
a by a Pad\'e-Borel-Leroy (PBL) resummation. In Ref. \cite{AharonyNeu} the 
resummation scheme of Ref. \cite{KoPa17} is used.
For a detailed discussion of these analyses, we refer the 
reader to the original work.

In Ref. \cite{DBinder} a large $N$-expansion around the decoupled Ising 
fixed point has been performed on the basis of the CB results for the 
Ising universality class,
see Ref. \cite{Simmons-Duffin:2016wlq} and references therein.
The results for critical exponents obtained in Ref. \cite{DBinder} are
\begin{equation}
\label{etaLN}
\eta = 
0.03629 + \frac{0.001232}{N} + O(N^{-2})
\end{equation}
and
\begin{equation}
\label{deltaLN}
\Delta^*_{\epsilon}  = 1.5874 + \frac{0.0796}{N} + O(N^{-2}) \;,
\end{equation}
where $\nu=1/(d-\Delta^*_{\epsilon})$.
In tables \ref{fieldtheoryN3} and \ref{fieldtheoryN4} we give
the numbers obtained from 
Eqs.~(\ref{etaLN},\ref{deltaLN}) by inserting $N=3$ and $4$, respectively.
Finally we give the results obtained in the present work.

Let us first discuss the numbers for $N=3$  summarized in 
table \ref{fieldtheoryN3}.
The estimates of $\nu$ obtained from the $\epsilon$-expansion by 
different authors are consistent. However they are too small compared 
with our result. They differ from our result by more than the error
that is quoted. The estimates of $\nu$ obtained from the perturbative series in
$d=3$ fixed are larger than those obtained from the $\epsilon$-expansion.
Still they are too small compared with ours. 
The estimate obtained from the large $N$-expansion is larger than ours.
But one should note that the deviation is of similar size as that for 
the $\epsilon$-expansion, which is quite remarkable given the small value
of $N$. 

In the case of $\eta$ we find that the estimates obtained from the 
analysis of the $\epsilon$-expansion are consistent with ours, while 
those obtained from the perturbative series in $d=3$ fixed are smaller
and the estimate of the error is smaller than the difference.

The  results for the correction exponents $\omega_1$
and $\omega_2$ obtained by different authors are essentially consistent.
Within errors $\omega_1$ of the cubic fixed point is the same as 
$\omega$ of the $O(3)$-invariant fixed point.  For $\omega_1$ we have no direct
numerical estimate. 
Our estimate of $\omega_2$ is larger than those obtained by field  theoretic
methods.

In table \ref{fieldtheoryO3} we have selected a few results for the 
critical exponents for the $O(3)$-invariant fixed point.  
At the level of the accuracy obtained by field theoretic methods, the 
estimates for the critical exponents for the cubic and the 
$O(N)$-invariant fixed point can not be discriminated for $N=3$. 

It is an interesting idea, to directly aim at 
the difference between the values of critical exponents for the 
$O(3)$-symmetric and the cubic fixed point. Given the fact that the two
fixed points are close in coupling space, one might hope that systematic
errors of the calculation are more or less the same and cancel when the 
difference is taken.

Such an analysis of the perturbative expansion in $d=3$ fixed is given
in the Appendix of Ref. \cite{Calabrese}. 
The authors find
\begin{eqnarray}
 \nu_{cubic} - \nu_{O(3)} &=& -0.0003(3)  \; ,  \\
\eta_{cubic} - \eta_{O(3)} &=& -0.0001(1) \; ,  \\
\gamma_{cubic} -\gamma_{O(3)} &=& -0.0005(7) \; .
\end{eqnarray}
Our estimate $\nu_{cubic} - \nu_{O(3)} = -0.00061(10)$, Eq.~(\ref{nudiff}),
is consistent with that of Ref. \cite{Calabrese}. In the case of 
$-0.00001 \lessapprox \eta_{cubic} - \eta_{O(3)} \lessapprox 0.00007$,
Eq.~(\ref{etadiff}), we favor the opposite sign as the authors
of Ref. \cite{Calabrese}.

Starting from the 6-loop $\epsilon$-expansion, 
the authors of Ref. \cite{AharonyNeu} perform an expansion of the RG-flow
around the $O(3)$-symmetric fixed point to second order.
Furthermore the authors have computed
effective critical exponents, depending on the parameters of this RG-flow, see
eqs.~(14) of Ref. \cite{AharonyNeu}. Plugging in the values of the parameters
for the cubic fixed point, the authors get $\gamma_{cubic}=1.3849(61)$ and
$\beta_{cubic}=0.3663(21)$. These values are virtually identical with 
$\gamma_{O(3)} =1.385(4)$ and $\beta_{O(3)}=0.3663(12)$  obtained in 
Ref. \cite{KoPa17} by using the same resummation scheme. Using the information
given by the authors it is hard to estimate the error of the difference, which
might be much smaller than the naively propagated one.

One also should note the discussion of section 5 of Ref. \cite{CB_O3}.
To leading order, the deviation of the exponents of the cubic fixed point
from those of the $O(3)$-invariant one is proportional to $Y_4$ and 
the coefficient is given by structure constants of the $O(3)$-invariant
fixed point. For $\nu$ and $\eta$, these coefficients vanish.

There have also been attempts to isolate the cubic fixed point for $N=3$ by 
using the CB method \cite{Ster18,KoSt19,KoSt20}.  However the candidate 
that is found, has critical exponents and a correction exponent very 
different from those discussed here.  

\begin{table}
\caption{\sl \label{fieldtheoryN3}
Estimates of the exponents $\nu$, $\eta$, and $\gamma$ and the correction 
exponents $\omega_1$ and $\omega_2$ for the cubic fixed point for $d=3$ and
$N=3$.  Aharony et al. \cite{AharonyNeu} only quote the result for 
the exponents $\beta$ and $\gamma$ (see their table II). They give
$\beta=0.3669(12)$. Inserting
our results for $\nu$ and $\eta$, we arrive at 
$\beta=\frac{\nu}{2} (d -2 + \eta) = 0.3690(2)$.  $^*$ indicates that 
the Monte Carlo result for $\gamma$ is obtained by inserting our numerical
estimates of $\nu$ and $\eta$ into $\gamma=\nu (2-\eta)$. For a discussion 
see the text.
}
\begin{center}
\begin{tabular}{|c|c|l|l|l|l|l|}
\hline
Ref. & method & \mc{1}{|c|}{$\omega_1$}  &  \mc{1}{c|}{$\omega_2$} &  
\mc{1}{c|}{$\nu$}   &  \mc{1}{c|}{$\eta$} &  \mc{1}{c|}{$\gamma$} \\
\hline
\cite{Mud98}&5-loop $\epsilon$-exp&  &   & 0.6997(24) & 0.0375(5) &1.3746(20)\\
\cite{Carmona}&5-loop $\epsilon$-exp&0.799(14)& 0.006(4)& 0.701(4)&0.0374(22)
 & 1.377(6) \\
\cite{Carmona}&6-loop $d=3$ fix& 0.781(4) & 0.010(4) & 0.706(6) & 0.0333(26) & 1.390(12) \\
\cite{Varna00} & 6-loop $d=3$ fix & 0.7833(54)&0.0109(32)&0.7040(40)&0.0327(20)  &   1.3850(50) \\
\cite{Folk00}& 6-loop $d=3$ fix & 0.777(9) &        &   0.705(1) &   & 1.387(1)\\
\cite{epsilon6} &6-loop $\epsilon$-exp, PBL & 0.799(4)  &  0.005(5)  & 0.700(8) &  0.036(3)  &   1.368(12) \\ 
\cite{epsilon6} &6-loop $\epsilon$-exp, Pad\'e & 0.78(11)  &  0.008(38) & 0.703(5) &  0.038(4)  &   1.379(8) \\ 
\cite{AharonyNeu}&6-loop $\epsilon$-exp &  &&        &       & 1.387(9) \\
\cite{DBinder} &Large $N$  &     &      & 0.7215   &  0.03671  & \\
this work     & MC  &     & 0.0133(8) & 0.7111(3)&  0.03782(13)   & 1.3953(6)$^*$ \\
\hline
\end{tabular}
\end{center}
\end{table}

\begin{table}
\caption{\sl \label{fieldtheoryO3}
Estimates of the exponents $\nu$, $\eta$,  $\gamma$, and $Y_4$
and the correction exponent $\omega$ for the Heisenberg fixed point for 
$d=3$ and $N=3$. $^*$ indicates that in the case of Ref. \cite{myIco}, 
we computed 
$\gamma=(2-\eta) \nu$ by using the values given for $\nu$ and $\eta$.
For a discussion
see the text.
}
\begin{center}
\begin{tabular}{|c|c|l|l|l|l|l|}
\hline
Ref. & method & \mc{1}{c|}{$\omega$}  &  \mc{1}{c|}{$Y_4$} &
\mc{1}{c|}{$\nu$}  & \mc{1}{c|}{$\eta$} &  \mc{1}{c|}{$\gamma$} \\
\hline
\cite{GuZi98} & 5-loop $\epsilon$-exp &0.794(18)& & 0.7045(55) &0.0375(45)&1.3820(90) \\
\cite{GuZi98} & $d=3$  & 0.782(13) &  &0.7073(35)& 0.0355(25) &1.3895(50) \\
\cite{KoPa17} & 6-loop $\epsilon$-exp &0.795(7) &      & 0.7059(20) & 0.0378(5) & \\
\cite{Carmona}& 5-loop $\epsilon$-exp & & 0.003(4) &            &   & \\
\cite{Carmona}&6-loop $d=3$ & & 0.013(6) &            &   & \\
\cite{AharonyNeu}  &6-loop $\epsilon$-exp &0.7967(57)  & 0.0083(15)&   &   & \\
 \cite{CB_O3} & CB &    & $>$0.00944    &   0.71169(30)  &  0.037884(102)&  \\
\cite{O234}   & MC &           & 0.013(4)&  & & \\ 
 \cite{myIco} & MC & 0.759(2)  &     &   0.71164(10)  &  0.03784(5)   &1.39635(20)$^*$   \\
this work & MC, Sec. \ref{O3exponents}  &    & 0.0143(9) &     &   & \\
this work & MC, Sec. \ref{UCflow}  &    & 0.0142(6) &     &   & \\
\hline
\end{tabular}
\end{center}
\end{table}

\begin{table}
\caption{\sl \label{fieldtheoryN4}
Estimates of the exponents $\nu$, $\eta$,  and $\gamma$ and the 
correction exponents $\omega_1$ and $\omega_2$ for the cubic fixed point 
for $d=3$ and $N=4$.
 $^*$ indicates that
the Monte Carlo result for $\gamma$ is obtained by inserting our numerical
estimates of $\nu$ and $\eta$ into $\gamma=\nu (2-\eta)$. For a discussion
see the text.
}
\begin{center}
\begin{tabular}{|c|c|l|l|l|l|l|}
\hline
Ref. & method & \mc{1}{c|}{$\omega_1$}  &  \mc{1}{c|}{$\omega_2$} &
\mc{1}{c|}{$\nu$}  & \mc{1}{c|}{$\eta$} &  \mc{1}{c|}{$\gamma$} \\
\hline
\cite{Mud98}&5-loop $\epsilon$-exp&  &  &0.7225(22) & 0.0365(5)&1.4208(30)\\
\cite{Carmona}&5-loop $\epsilon$-exp&0.790(8)&0.078(4)&0.723(4)&0.0357(18) &
1.419(6) \\
\cite{Carmona}&6-loop $d=3$ fix&0.781(44)&0.076(40)&0.714(8)&0.0316(22)& 
1.405(10) \\
\cite{Varna00}&6-loop $d=3$ fix & 0.7887(90) & 0.0740(65) &0.7150(50)&
 0.0316(25) & 1.4074(30) \\
\cite{Folk00}& 6-loop $d=3$ fix &0.777(2) &  &0.719(2) &   & 1.416(4) \\
\cite{DBinder} &Large $N$  &     &      &0.7180    & 0.03661   & \\
this work &MC&  0.763(24)&0.082(5)&0.7202(7) &0.0371(2) &1.4137(14)$^*$\\
\hline
\end{tabular} 
\end{center}
\end{table}
Let us discuss the results for $N=4$ summarized in table \ref{fieldtheoryN4}.
 Here we see that the estimates 
of $\nu$ obtained by the various authors are consistent with our result.
The estimate obtained by the large $N$-expansion is slightly smaller
than ours. Note that for $N=3$ it is bigger and the deviation is roughly 
by a factor of 4 larger than for $N=4$.  It is plausible that for $N \ge 5$
the large $N$-expansion expansion gives very accurate results and might 
serve as benchmark for the analysis of the $\epsilon$-expansion or the 
perturbative expansion in $d=3$ fixed.   In the case of the exponent 
$\eta$ the findings are similar to $N=3$. The results obtained from the 
$\epsilon$-expansion are consistent with ours, while those obtained 
from the perturbative expansion in $d=3$ fixed are too small. The estimate
obtained from the large $N$-expansion is slightly smaller than ours.
The deviation is much smaller than for $N=3$. 
It is plausible that for $N \ge 5$  the deviation of the large $N$-expansion
from the exact value is at most in the $5^{th}$ digit.

The estimates of $\omega_2$ are consistent among different authors and 
the field theoretic results are consistent with that obtained here. Our 
estimate of $\omega_1$ is smaller than that obtained by field theoretic
methods, which also holds in the case of $\omega$ for the $O(4)$-symmetric 
fixed point \cite{myON}. Within error bars our estimate of $\omega_1$
takes the same value as $\omega$ for the $O(4)$-symmetric 
fixed point \cite{myON}.

In contrast to $N=3$, truncating the expansion of the scaling fields
of the $O(N)$-invariant fixed point at second order is no good approximation.
This is seen for example by the fact that $Y_4$ and $\omega_2$ clearly 
differ.

\section{Summary and Conclusions}
\label{summary}
We have studied a $\phi^4$ model on the simple cubic lattice, where 
the reduced Hamiltonian, Eq.~(\ref{Hamiltonian}),  includes
a term that breaks $O(N)$ invariance and possesses only cubic symmetry.
It has two parameters $\lambda$ and $\mu$, where $\mu$ controls the 
breaking of the $O(N)$ invariance. Field theory predicts that for $N>N_c$
the perturbation of the $O(N)$-invariant fixed point is relevant,
where $N_c$ is slightly smaller than three. In fact
the recent conformal bootstrap study \cite{CB_O3} finds the rigorous
bound $Y_4> 3-2.99056$ for the RG-exponent of a cubic perturbation for $N=3$. 
Depending on the sign of the parameter $\mu$, the system should undergo
a first order phase transition or a continuous transition governed by the 
cubic fixed point. For a recent discussion of the implications
for structural transition in perovskites see Ref. \cite{AharonyNeu}.

For $N=4$ the cubic fixed point is well separated from the $O(4)$-symmetric one.
Using a finite size scaling analysis of dimensionless quantities such
as the Binder cumulant $U_4$, we determine the improved model,
characterized by $(\lambda,\mu)^*$, where the two leading corrections 
to scaling vanish. 
In order to monitor the violation of the $O(N)$ symmetry the 
cumulant $U_C$, Eq.~(\ref{UCdef}), is introduced. It vanishes for an 
$O(N)$-symmetric distribution of the order parameter.
At $(\lambda,\mu)^*$, we determine the critical exponents $\nu$ and $\eta$
by using standard finite size scaling methods. For $N=4$ these are clearly
different from those of $O(4)$-symmetric systems. For the correction exponents
we obtain $\omega_1=0.763(24)$ and $\omega_2=0.082(5)$ for  $N=4$.
One should note that in order to reduce the effect of corrections 
proportional to $L^{-\omega_2}$ for example by half, 
one has to increase the linear lattice 
size by the factor $2^{1/\omega_2}  \approx 4700$. It is clear that
in a Monte Carlo study of lattice models, we can not approach the cubic
fixed point by just increasing the linear lattice size $L$. It is mandatory
to eliminate corrections proportional to $L^{-\omega_2}$ by a proper choice 
of the parameters!  This is even more the case for $N=3$, where we find 
$\omega_2=0.0133(8)$.

In the experimentally relevant case $N=3$, the cubic fixed point is close
to the $O(3)$-invariant one. This is related to the fact that the correction 
and RG exponents  $\omega_2 \approx Y_4 = 0.0143(9)$ have a small modulus.
This also implies that there is a slow RG-flow along a line in coupling space.
In order to analyze the behavior of dimensionless quantities we use 
Ans\"atze that are approximately valid in a region of the parameter space that
includes both the $O(3)$-symmetric and the cubic improved models.
This allows us to determine a line $\lambda^*(\mu)$ in the $(\lambda,\mu)$
plane, onto which the RG-flow rapidly collapses. 

In order to study the flow of the symmetry breaking perturbation, we focus
on the dimensionless quantity $U_C$. Based on the RG-flow equation 
to second order, Eq.~(\ref{RGeq}), we obtain an Ansatz for $U_C$ that is a good 
approximation
in a region of the parameter space that includes both improved models.
We obtain the estimate  $(\lambda,\mu)^* =  (4.99(11),0.28(2))$,
characterizing the 
improved model for the cubic fixed point. 
Estimates of the exponents $\nu$ and $\eta$ of the cubic
fixed point are obtained by analyzing the slopes of dimensionless quantities
and the magnetic susceptibility at $(\lambda,\mu)=(5.0,0.3)$ and values
close by. It turns out that the estimate of $\eta$ is the same as that 
for the $O(3)$-symmetric fixed point within errors. In the case of the 
exponent of the correlation length the estimate $\nu_{cubic}=0.7111(3)$ 
obtained for 
the cubic fixed point is only slightly smaller than that for the 
$O(3)$-symmetric one $\nu_{O(3)}=0.71164(10)$ \cite{myIco}. Since we have
estimated the error conservatively here, we consider the difference 
as significant.

In Sec. \ref{UCflow} we go beyond the second order approximation of the 
RG-flow. In the second order approximation $Y_4=\omega_2$, while in 
Sec. \ref{UCflow} we find $0 \lessapprox Y_4-\omega_2 \lessapprox 0.0015$. 

In Sec.~(\ref{ratio_slope}) we analyze ratios of magnetic susceptibilities
and slopes of dimensionless quantities to get estimates of the differences
of the critical exponents for the cubic and the $O(3)$-invariant fixed point.
The idea is that subleading corrections approximately cancel, and the 
systematic error is reduced in the difference. In fact, we arrive 
at $-0.00001  \lessapprox \eta_{cubic} - \eta_{O(3)} \lessapprox 0.00007$
and $\nu_{Cubic}-\nu_{O(3)} =-0.00061(10)$.

The results of the present work can be improved by simply increasing the 
statistics and moderately increasing the linear lattices sizes. Beyond
that we would like to extend the study for $N=3$ in the following directions:

\begin{itemize}
\item 
Study $|\mu| > 1$. In particular we would like to
extend the flow equation for $\bar{U}_C$ discussed in Sec. \ref{UCflow}
to larger values of $|\bar{U}_C|$. On the 
one hand we like to extend the range up to the decoupled Ising fixed
point and on the other hand we like to see clear signs of the 
first order transition in the simulation.

\item
Here we studied finite systems at criticality. It would be interesting
to study the case $\xi \ll L$ that approximates the thermodynamic limit
in the phases. One could compute universal amplitude ratios that can 
be compared with results obtained in experiments.

\item
Extending the calculation of RG-exponents to a larger set of operators. 
In section 5 of Ref. 
\cite{CB_O3} it is discussed that for example the RG-exponent
of the rank-2 symmetric 
tensor should have a contribution at leading order in $Y_4$, in contrast 
to the singlet and vector operators studied here.
\end{itemize}

\section{Acknowledgement}
This work was supported by the Deutsche Forschungsgemeinschaft (DFG) under
the grant  HA 3150/5-3.

\end{document}